\documentclass[aps,twocolumn,secnumarabic,nobalancelastpage,amsmath,amssymb,nofootinbib,floatfix, superscriptaddress]{revtex4-1}

\usepackage{hyperref}  
\hypersetup{colorlinks}
\usepackage{graphicx}      % tools for importing graphics
\usepackage{bm}            % special bold-math package. usge: \bm{mathsymbol}
\usepackage{caption}
\usepackage{subcaption}
\usepackage{overpic}
\usepackage[separate-uncertainty=true,range-phrase=\:--\:]{siunitx}  % For formatting units in text

\usepackage{subfiles} % to include supplement at end of the document

\begin{document}
\title{Magneto-Optical Properties of InSb for Infrared Spectral Filtering}
\author{Nolan Peard}
\date{\today}
\affiliation{Department of Applied Physics, Stanford University, Stanford, CA 94205, USA}
\affiliation{The Charles Stark Draper Laboratory, Inc., Cambridge, MA 02139, USA}
\author{Dennis Callahan}
\author{Joy C. Perkinson}
\affiliation{The Charles Stark Draper Laboratory, Inc., Cambridge, MA 02139, USA}
\author{Qingyang Du}
\affiliation{Department of Materials Science and Engineering, Massachusetts Institute of Technology, Cambridge, MA 02139, USA}
\author{Neil S. Patel}
\affiliation{The Charles Stark Draper Laboratory, Inc., Cambridge, MA 02139, USA}
\author{Takian Fakhrul}
\affiliation{Department of Materials Science and Engineering, Massachusetts Institute of Technology, Cambridge, MA 02139, USA}
\author{John LeBlanc}
\affiliation{The Charles Stark Draper Laboratory, Inc., Cambridge, MA 02139, USA}
\author{Caroline A. Ross}
\email{caross@mit.edu}
\author{Juejun Hu}
\email{hujuejun@mit.edu}
\affiliation{Department of Materials Science and Engineering, Massachusetts Institute of Technology, Cambridge, MA 02139, USA}
\author{Christine Y. Wang}
\email{cywang@draper.com}
\affiliation{The Charles Stark Draper Laboratory, Inc., Cambridge, MA 02139, USA}

\begin{abstract}
We present measurements of the Faraday effect in n-type InSb. The Verdet coefficient was determined for a range of carrier concentrations near \SI{e17}{\centi\metre^{-3}} in the $\lambda = $ \SIrange{8}{12}{\micro\metre} long-wave infrared regime. The absorption coefficient was measured and a figure of merit calculated for each sample. From these measurements, we calculated the carrier effective mass and illustrate the variation of the figure of merit with wavelength. A method for creating a tunable bandpass filter via the Faraday rotation is discussed along with preliminary results from a prototype device. 
\end{abstract}

%Absent from this paper: Complete measurements of the refractive index via ellipsometry, study of property change with temperature

\maketitle

\section{Introduction}
Faraday rotation of the polarization state of light is one of the most well-known non-reciprocal processes in physics. It is a result of circular birefringence induced in materials with strong magneto-optical coupling by an applied magnetic field. Materials exhibiting large Faraday rotation, in combination with polarization optics, are used to break time-reversal symmetry for light propagation, allowing the use of these materials as isolators in optical communications to protect light sources and detectors from back-reflections in the optical path. The Faraday effect has been demonstrated in all classical states of matter, but only solid-state materials find widespread application. 

% Paragraph about hyperspectral imaging applications here?

Crystalline semiconductors with high carrier mobility are ideal magneto-optical materials due to the high density of free carriers with low effective mass and long mean free time. Such materials have been used for decades to implement large Verdet coefficients ($V = \Delta \theta/|B|L$) - the constant of proportionality between the change in polarization angle ($\Delta \theta$), magnitude of the applied magnetic field ($|B|$), and material thickness ($L$) \cite{Ukhanov_1974}. In the long-wave infrared (LWIR) range of interest for thermal imaging ($\lambda = $ \SIrange{8}{12}{\micro\metre}), the best-performing Faraday rotators are doped semiconductors where the magneto-optical coupling is dominated by the free carrier effect \cite{Boord_1974, Hilico_2011, Tomasetta_1979, Smith_1959}. Taking the Drude model for a free-electron gas, it can be shown that the Verdet coefficient varies linearly with the doping concentration ($N$) and inversely with the square of the electron effective mass ($m^*$) as shown in Equation \ref{eq:Verdet}.

\begin{equation} V =  \frac{ \mu_o N q^3 \lambda^2}{8 \pi^2  {m^*}^2  n c } \label{eq:Verdet} \end{equation}
where $\mu_0$ is the vacuum permeability, $q$ is the carrier charge ($-1$ for electrons), $n$ is the refractive index, and $c$ is the speed of light.

The III-V semiconductor InSb has among the lowest electron effective masses of any known material ($m^* \approx \SI{0.014}{\electronmass}$ for intrinsic InSb) as well as good transparency below the band edge absorption (\SI{0.17}{\electronvolt}), making it a promising material for magneto-optical devices at long wavelengths \cite{Boer_Pohl_2018, Smith_1959}. In addition to polarization rotation efficiency, low attenuation is equally important for infrared optical devices and is quantified by the absorption coefficient, $\alpha$. For InSb, this is again derived in the Drude free-electron gas limit where scattering events are temporally-separated by mean free time $\tau$, resulting in Equation \ref{eq:Absorb}.

\begin{equation} \alpha =  \frac{ \mu_o N q^2 \lambda^2}{4 \pi^2  m^*  n \tau c } \label{eq:Absorb} \end{equation}

It is useful to define a figure of merit (FOM) that accounts for both rotation efficiency and optical loss

\begin{equation} \text{FOM} =  \frac{V}{\alpha} = \frac{q\tau}{2m^*} = \frac{\mu}{2} \label{eq:FOM} \end{equation}  where $\mu$ is the carrier mobility \cite{Hilico_2011, Boord_1974}. 

Materials with high free-carrier densities typically exhibit large Verdet coefficients across a spectral range defined by their transparency window with large $\lambda^2$ dispersion. At shorter wavelengths approaching the electronic band gap, free-carrier rotation is reduced and interband absorption dominates. Since interband rotation opposes free carrier rotation and is independent of doping, there will exist a doping region where rotation is very low and a single doping where it is zero \cite{Tomasetta_1979, Smith_1959, Lax_Nishina_1961}. Additionally, the Moss-Burstein effect will modify the optical band gap, altering the usable range of wavelengths. Because of this trade-off and the doping dependencies of the effective mass and scattering time, there exists an optimal doping that maximizes the FOM in the free carrier dominated range. Some of these dependencies can be easily modeled via the electronic band structure (e.g. Verdet coefficient versus doping), while others are more complicated (e.g. temperature dependencies or absorption versus doping) or may depend heavily on sample preparation and purity \cite{Boord_1974, Jensen_1971, Tomasetta_1979}. Therefore, an experimental effort is warranted to determine the optimal doping for maximizing the FOM. 

Our research demonstrates that the ideal carrier concentration in InSb at room temperature for applications in the LWIR past \SI{8}{\micro\metre} is \SIrange{3e17}{5e17}{\centi\metre^{-3}}, where all wavelengths in the range \SIrange{8}{12}{\micro\metre} have roughly the same FOM. Moreover, we illustrate that the Drude model of absorption and the FOM are incorrect for InSb since additional absorption mechanisms, notably polar optical scattering, are present. We expect that a clear understanding of the Verdet and absorption coefficients in InSb with different carrier concentrations and temperatures will prove important for the future design of magneto-optical elements for hyperspectral imaging and other infrared photonics applications \cite{Gupta_2008, Rogalski_2014}. As an example, we present preliminary results for a prototype magneto-optical device which utilizes the dispersion of the Verdet coefficient to implement a tunable bandpass filter.

\section{Materials and Methods}
\subsection{InSb Wafers and Two-Stage Filter}
Custom Te-doped InSb crystals were grown by WaferTech LLC and cut so as to give a different carrier concentration and crystal orientation in each sample. Natural dopant gradients in the as-grown boules gave carrier concentrations ranging from $1.5 \times 10^{17}$ to $6 \times 10^{17}$ carriers per cubic centimeter in different wafers. 3-inch wafers were double-side polished to provide smooth optical quality surfaces with minimal scattering and diced into \SI{1}{\centi\metre^2} samples.

The two-stage filter was constructed as shown in Figure \ref{fig:Filter_Schematic}: A wire-grid polarizer was sandwiched between two InSb chips (\SI{0.8}{\milli\metre} and \SI{1.6}{\milli\metre} thickness) with anti-reflection (AR) coatings. The three layers were bonded at the edges by UV-curing optical adhesive. Commercially-available proprietary AR coatings were used for constructing this prototype bandpass filter. The front and back polarizers were mounted on rotating mounts with angles adjusted relative to the wire-grid polarizer as described in Section \ref{sect:Device}.

\subsection{Carrier Concentration and Mobility}
Carrier concentration and mobility were measured via the Hall effect using the Van der Pauw method on a Model 8404 AC/DC Hall Effect Measurement System by Lake Shore Cryotronics, Inc. \SI{100}{\nano\metre} gold electrical contacts were deposited on the corners of each piece using an AJA electron beam evaporator. The carrier concentration was calculated via \begin{equation} N = \frac{I|B|}{qL|V_H|} \end{equation} where $N$ is the bulk carrier concentration, $I$ denotes the electric current flowing through the sample, $|B|$ represents applied magnetic induction, $q$ is the unit charge, $L$ is the sample thickness, and $V_H$ is the average Hall voltage. 

%During measurements, samples were placed in a magnetic field with electric probes in contact with the 4 gold pads. A constant electric current is flowed between one diagonal pair while the other pair monitors the Hall voltage. The Hall voltage was recorded multiple times and averaged by reversing the magnetic field and reversing the electric current flow direction. The carrier concentration can be calculated via \begin{equation} N = \frac{IB}{qt|V_H|} \end{equation} where $N$ is the bulk carrier concentration, $I$ denotes the electric current flowing through the sample, $B$ represents applied magnetic induction, $q$ is the unit charge, $t$ is the sample thickness, and $V_H$ is the average Hall voltage.

The measurement of the Hall effect also yielded the carrier mobility, $\mu$. Combined with the carrier effective mass, it is possible to calculate the Drude mean free time via the relation \begin{equation} \mu = \frac{q \tau}{m^*} \end{equation} where $\tau$ is the mean free time. The effective mass ($m^*$) for n-type InSb of similar doping levels is found in the literature to be around \SI{0.014}{\electronmass} (intrinsic), but this value is dependent on the dopant concentration \cite{Spitzer_Fan_1957}. Measurement of the Verdet coefficient across different samples allowed us to determine the variation of the effective mass with dopant concentration.

%We can calculate the cyclotron frequency from our measured results for the Verdet coefficient to find the effective mass as a function of applied magnetic field via the relation \begin{equation} m = \frac{eB}{\omega_c \gamma} \end{equation}  where $\gamma$ is the relativistic correction to the cyclotron frequency. It is thus possible to calculate the mean electron scattering time and effective mass as a function of applied magnetic field. 

\subsection{Verdet Coefficient}
\begin{figure}
\includegraphics[width=\columnwidth]{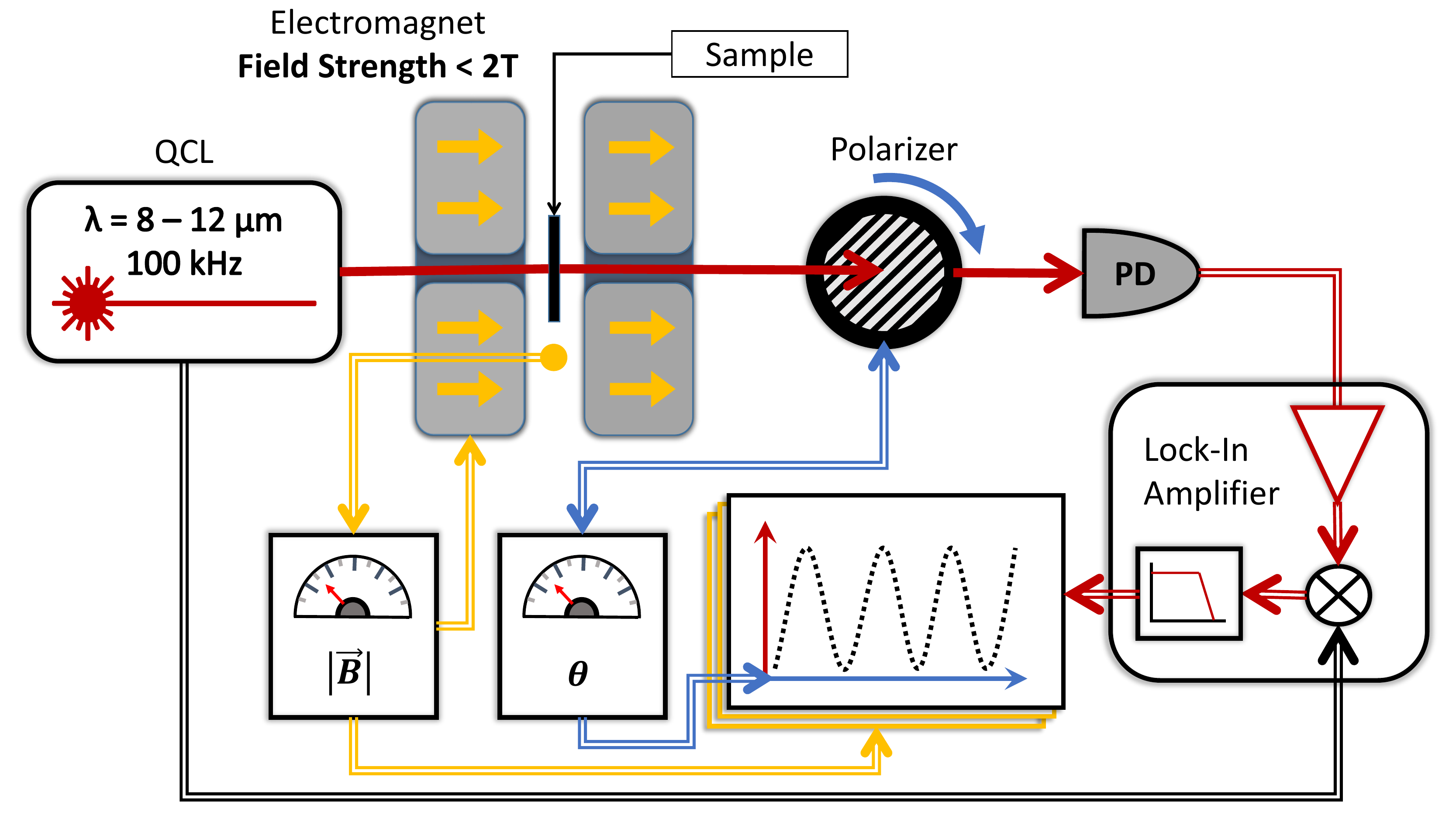}
\caption{Experimental setup and signal chain for Verdet coefficient measurements of InSb.  See text for details. \label{fig:VerdetSetup}}
\end{figure}

A schematic of the experimental setup is shown in Figure \ref{fig:VerdetSetup}. Samples were mounted between the poles of a GMW Associates Model 5405 Electromagnet. A tunable, linearly-polarized Block Engineering LaserTune quantum-cascade laser (QCL) was aimed through an aperture in the magnet. Infrared radiation from the QCL propagated parallel to the magnet field lines, interacted with the material, and exited the magnet at the opposite end. The polarization state of the light after interaction was measured with a rotating polarizer and photodetector (Block Engineering External IR Detector Module); the intensity at the detector varied with the relative angle of the polarizer. A lock-in amplifier (Stanford Research Systems SR844) was used to demodulate the pulsed laser intensity at the detector using the QCL controller as an external trigger. Multiple shots of the QCL were measured for each polarizer angle to determine the average intensity; the polarizer was stepped by a ThorLabs PRM1Z8 precision rotation stage for a full \SI{360}{\degree} rotation in increments of \SIrange{5}{10}{\degree}. The change in polarization angle due to the Faraday effect was observed by measuring the change in the absolute polarization of the escaping light as the strength of the magnetic field was varied in the material. The magnetic field strength was modulated by manipulating the current in the electromagnet and measured near the sample location using a Senis Hall I1A probe from GMW Associates; the maximum achievable field in the setup was approximately \SI{1.7}{\tesla} since the magnet poles could not be fully closed with the sample stage in place. The intrinsic background rotation of the experimental setup was determined by following the same procedure without a sample present in the chamber and measured to be less than \SI{0.25}{\degree \tesla^{-1}} for all wavelengths.

%All samples were measured at vacuum (9-100 mTorr) to reduce the probability of adsorbed molecules on the sample surfaces and were thermally-anchored to the temperature-controlled stage using Apiezon N Grease and a pressure clamp (temperature calibration curves may be found in the Supporting Information). The intrinsic background rotation of the experimental setup was determined by following the same procedure without a sample present in the chamber and measured to be less than 0.25 deg/T for all wavelengths. \par

The transmission intensity data were analyzed by fitting a surface of the form $I(\phi, |B|) = A \cos^2\left(\frac{\pi}{180} (\phi + \Delta \theta + \delta) \right) + C$ where $A$, $C$, and $\delta$ are constants, $|B|$ is the magnetic field strength, and $L$ is the thickness of the material. Assuming that the Faraday rotation was linear with the magnetic field, the Verdet coefficient is obtained from $\Delta \theta = V|B|L$. Measurements of the thickness $L$ were determined to better than \SI{10}{\micro\metre} precision for each sample using a micrometer. All data fitting was performed using orthogonal distance regression (ODR), accounting for the experimental uncertainty in both the dependent and independent variables. An example of the collected experimental data and accompanying fit of the model may be seen in the Supporting Information. The FOM was calculated from the Verdet coefficient and absorption coefficients extracted from FTIR measurements of each sample. Precision of Faraday rotation measurements were on the order of \SI{1}{\degree} for each sample and wavelength, such that the uncertainty in the Verdet coefficient was dominated by precision of the sample thickness measurements ($\approx \SI{3}{\micro\metre}$).

\subsection{Absorption Coefficient}
%Infrared transmission spectra were acquired using a Bruker Hyperion 2000 FTIR microscope. Background spectra were acquired before measuring each sample and multiple scans were performed to calculate averaged spectra. Wafer-scale FTIR maps were acquired using a Pike Technologies MappIR accessory to estimate the wafer doping uniformity. Both the MappIR and microscope accessories utilized a Bruker Tensor II FTIR Spectrometer.

Infrared transmission spectra were measured using a Bruker Tensor II FTIR Spectrometer. Samples were placed in the focus of the broadband IR beam and apertures applied to ensure use of near-collimated light at normal incidence. Background spectra were acquired before each measurement and multiple scans performed to compute averaged spectra for each sample.

The transmission spectrum from each sample was used to estimate the absorption coefficient, $\alpha = \frac{4 \pi k}{\lambda}$, which, along with the Verdet coefficient, determined the FOM at each wavelength. $k$ is the imaginary part of the complex-valued refractive index. $\alpha$ is defined such that incident intensity ($I_0$) is attenuated to $I(L) = I_0 e^{-\alpha L}$, for each pass through the material. Accounting for multiple internal reflections via the Fresnel equations, the expression for $\alpha$ is \begin{equation} \alpha = \frac{1}{L} \ln \left(\frac{-T^2 + \sqrt{T^4+4 T_{\text{total}}^2 R^2}}{2 T_{\text{total}} R^2}\right) \end{equation} where $L$ is the thickness of the material. Since $T$ and $R$, the Fresnel coefficients for transmission and reflection on a single pass through the material, were determined from the real refractive index $n$, it was important to verify that $n$ has small dispersion ($\frac{dn}{d\lambda}$) at room temperature in the LWIR. Including the dispersion of $n$ produced only a small correction to the absorption results (see Supporting Information) and did not alter the conclusions of Section \ref{sect:Absorb_Results} which assumed a constant $n = 3.927$ \cite{Adachi_1989}.

\section{Results}
\subsection{Electrical Properties}
\begin{figure*}[h!]
    \begin{subfigure}{0.49\textwidth}
        \begin{overpic}[unit=1mm,scale=0.49]{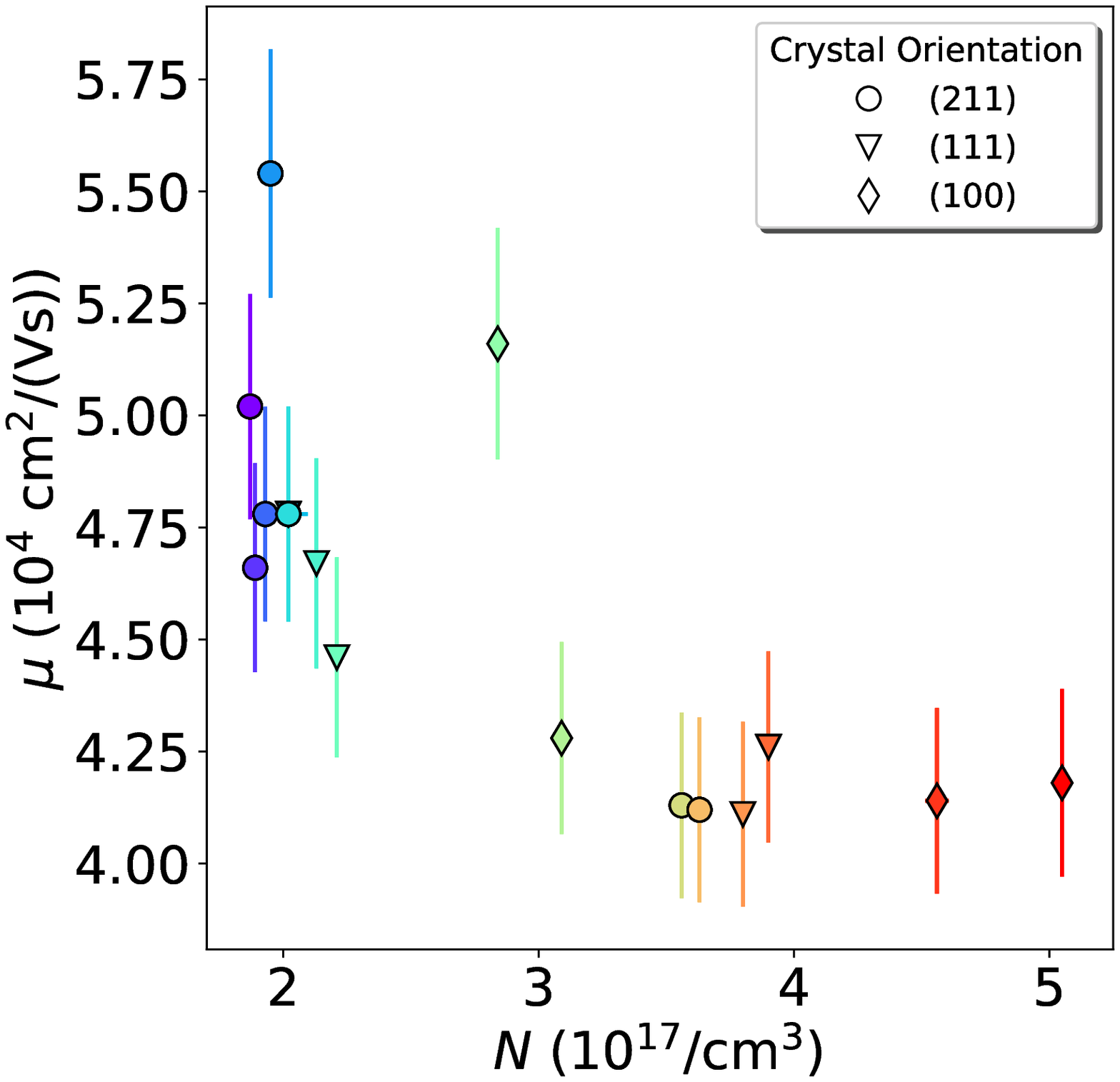}
        \put(20,20){\color{blue}\huge (a)}
        \end{overpic}
        \captionlistentry{}
        \label{fig:Electric_Mobility}
    %\includegraphics[width=\columnwidth]{figures/Electric_Mobility.png}
    %\caption{Carrier mobility as a function of carrier concentration with different crystal orientations measured using the Hall effect. Color corresponds to carrier concentration and marker shapes indicate the different crystal orientations. \label{fig:Electric_Mobility}}
    \end{subfigure}
    \hfill
    \begin{subfigure}{0.49\textwidth}
        \begin{overpic}[unit=1mm,scale=0.49]{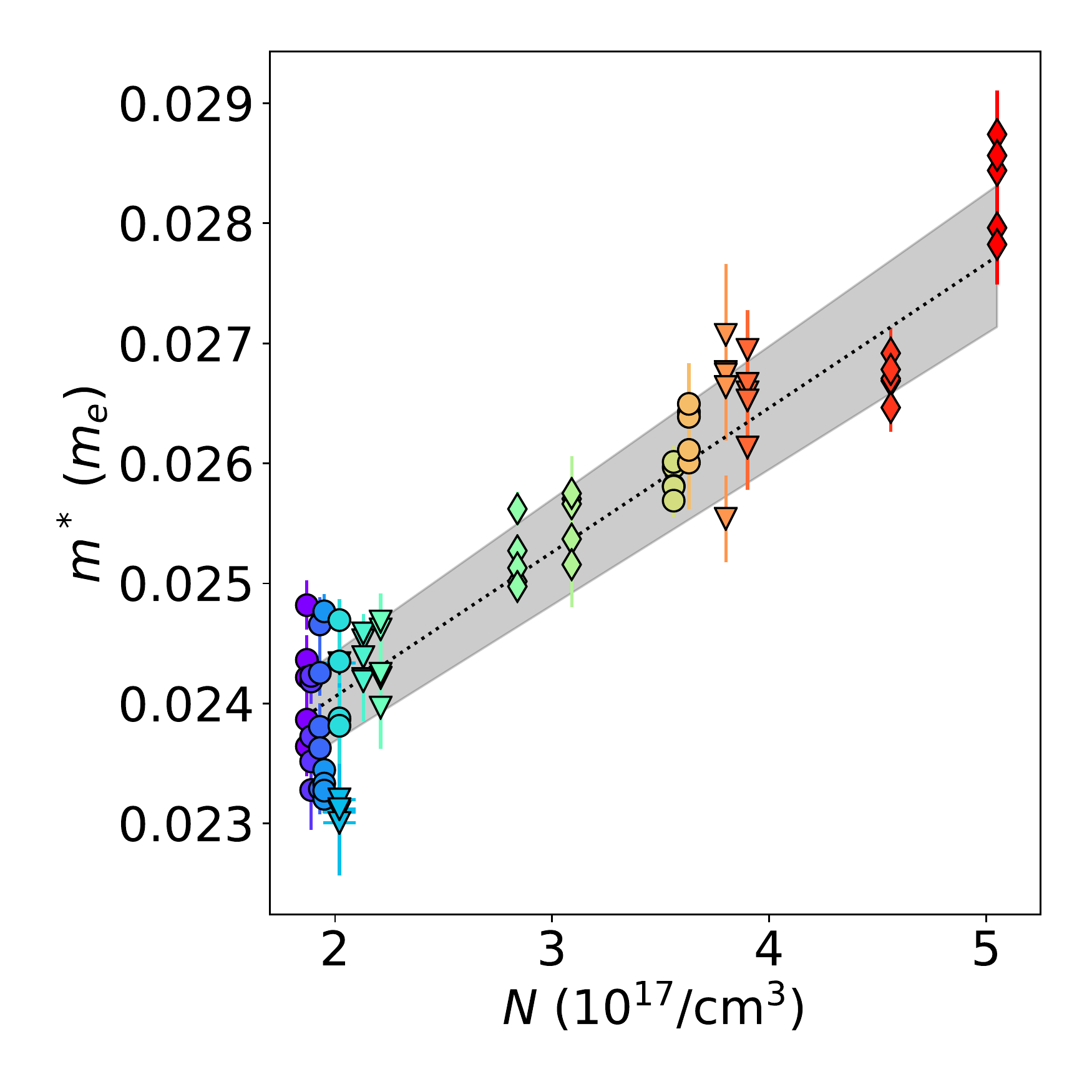}
        \put(26,87){\color{blue}\huge (b)}
        \end{overpic}
        \captionlistentry{}
        \label{fig:EffMass}
    %\includegraphics[width=\columnwidth]{figures/Fig_EffMass.png}
    %\caption{The effective mass calculated from our measured Verdet coefficients and Equation \ref{eq:Verdet} for each wavelength and dopant concentration measured. The data points are shown above with the color gradient corresponding to the dopant concentration. A linear trend is observed for this range of dopant concentrations and the line of best fit with error region is shown in black. \label{fig:EffMass}}
    \end{subfigure}
    \vspace{-5ex}
    \caption{ \ref{fig:Electric_Mobility} Carrier mobility as a function of carrier concentration with different crystal orientations measured using the Hall effect. Marker color scales with the carrier concentration on the horizontal axis and marker shapes indicate the different crystal orientations (consistent between all figures in this paper). \ref{fig:EffMass} The effective mass calculated from our measured Verdet coefficients and Equation \ref{eq:Verdet} for each wavelength and dopant concentration measured. A linear trend $(m^*(N) = (120.3 \pm 7.2) \times 10^{-5}N + (216.5 \pm2.3) \times 10^{-4})$ is observed for this range of dopant concentrations and the line of best fit with error region is shown in black. Vertically stacked data points are for different wavelengths in \SIrange{8}{12}{\micro\metre}.}
    
 \vspace*{\floatsep}
 
\begin{subfigure}{0.49\textwidth}
        \begin{overpic}[unit=1mm,scale=0.49]{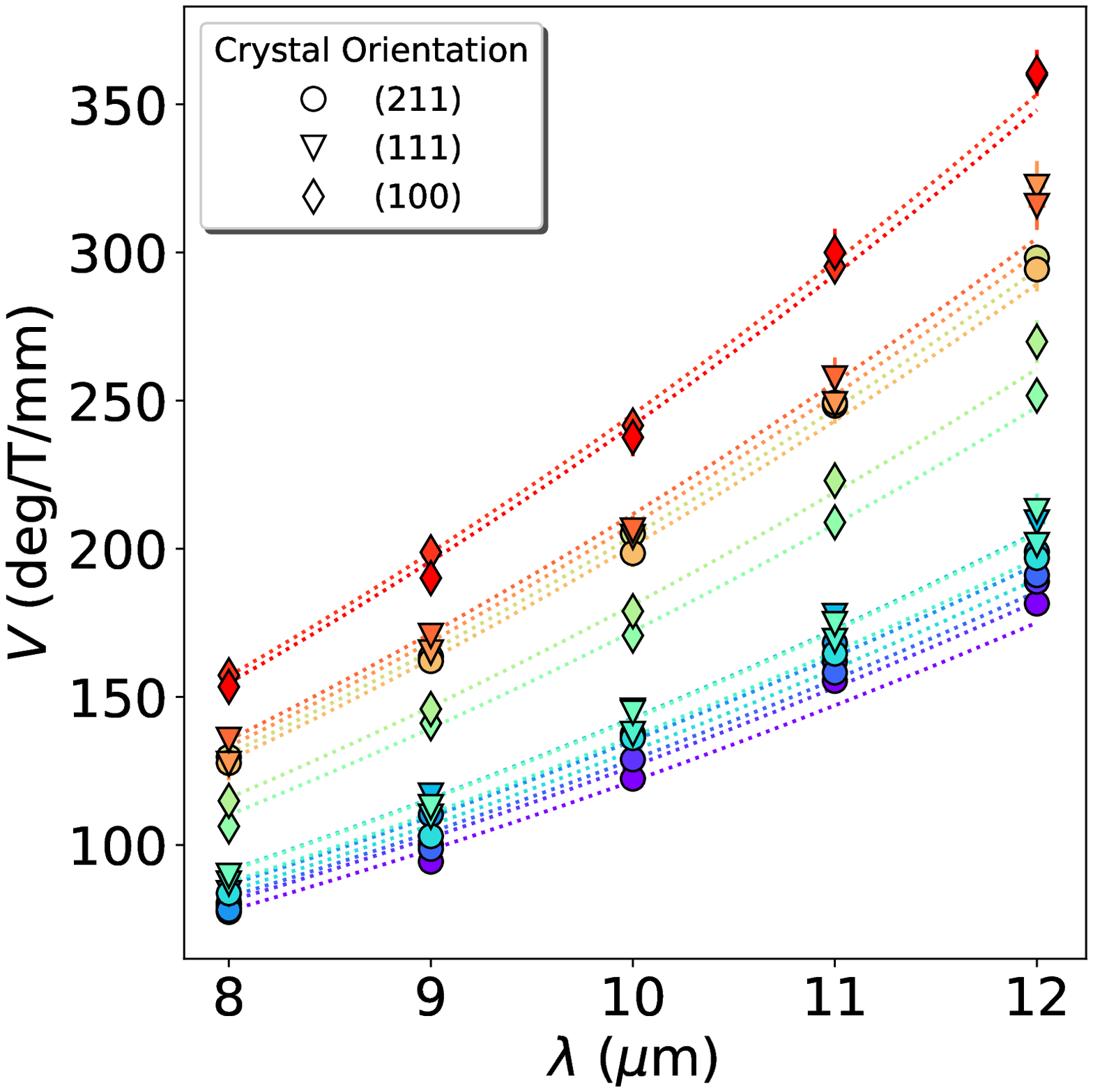}
        \put(85,20){\color{blue}\huge (a)}
        \end{overpic}
        \captionlistentry{}
        \label{fig:Verdet_Lambda}
\end{subfigure}
\hfill
\begin{subfigure}{0.49\textwidth}
        \begin{overpic}[unit=1mm,scale=0.49]{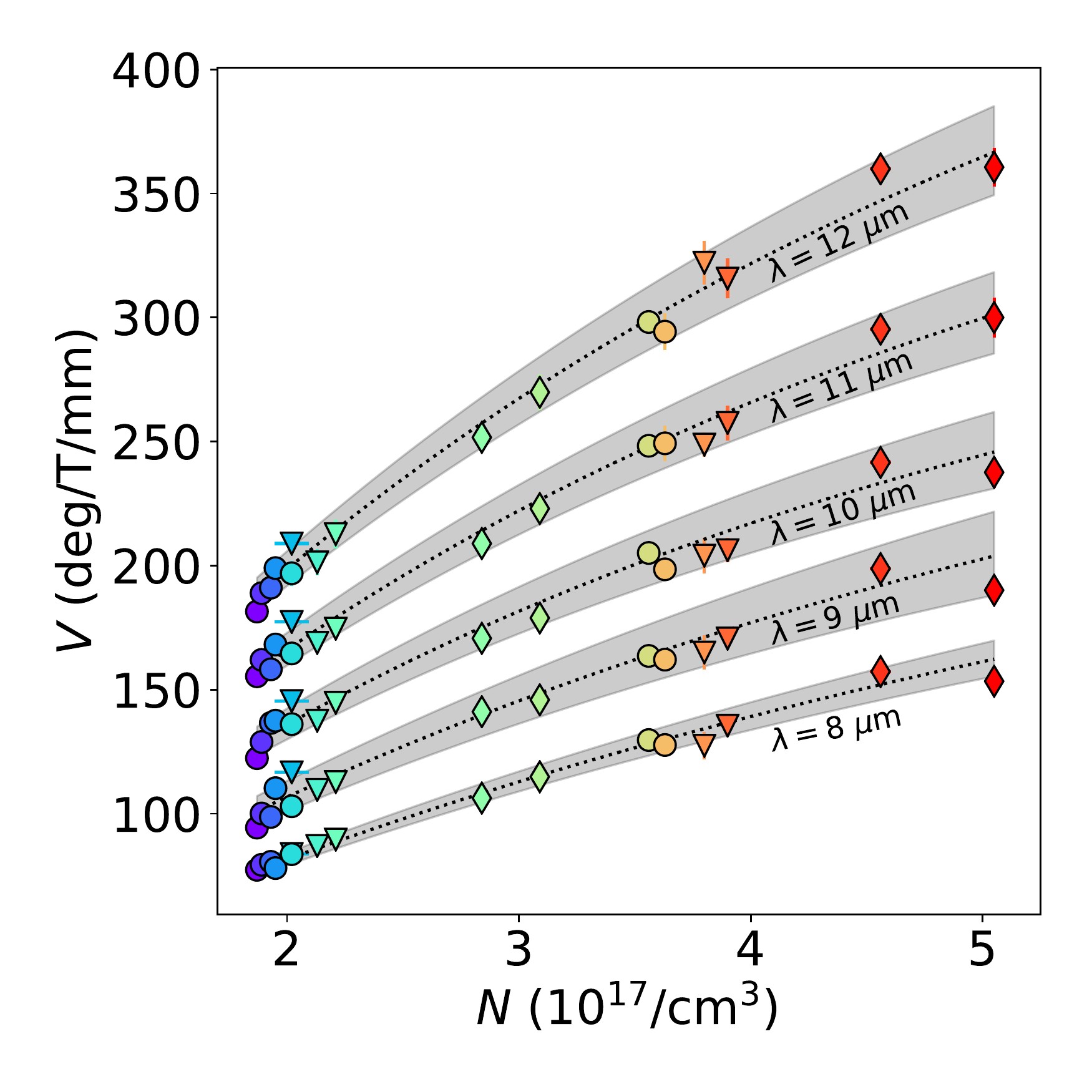}
        \put(85,20){\color{blue}\huge (b)}
        \end{overpic}
        \captionlistentry{}
        \label{fig:Verdet_Doping}
\end{subfigure}
\vspace{-5ex}
\caption{\ref{fig:Verdet_Lambda} Measurements of the Verdet coefficient at 8, 9, 10, 11, and 12 micron wavelengths in InSb were shown to roughly match a $\lambda^2$ dispersion as in Equation \ref{eq:Verdet} (colored dotted lines). \ref{fig:Verdet_Doping} The same Verdet coefficients are plotted against the measured carrier concentration of each sample. Here, the linear dependence of the Verdet coefficient on carrier concentration is modified by a linear correction to the effective mass which accounts for the dependence of the effective mass on carrier concentration (see Figure \ref{fig:EffMass}) and produces a concave curve for each wavelength (dotted black lines with error regions). In both figures, the marker color scales with the carrier concentration on the horizontal axis of \ref{fig:Verdet_Doping}. Where they are not visible, error bars have been eclipsed by their corresponding marker points. \label{fig:VerdetResults}}
\end{figure*}

InSb has a remarkably high mobility on the order of \SI{e4}{\centi\metre^2/{\volt \second}}. A high mobility ($\mu = \frac{q \tau}{m^*}$) implies a combination of long mean free time $\tau$ and small effective mass $m^*$ which determine the free-carrier lifetime and plasma frequency. Therefore, materials with high mobility should provide the strongest polarization rotation with little attenuation. Our results for measurements of the mobility in InSb may be seen in Figure \ref{fig:Electric_Mobility}. The mobility decreases at higher doping concentrations due to stronger scattering effects and increased effective mass. No clear differences between mobilities in samples with different crystal orientations are discernible. This is unsurprising considering the spherical isoenergy surface of InSb, which implies that the effective mass should be isotropic in the crystal momentum vector since the curvature of the conduction band is identical in each direction of reciprocal space \cite{Jung_1991, Kane_1957, Jensen_1973, Boer_Pohl_2018, Basov_1976}. Throughout this work, we observed no significant differences in optical, magneto-optical, or electrical properties of samples with different crystal orientations that could not be attributed to the non-identical carrier concentrations in those samples.

\subsection{Verdet Coefficient}
Our results for the dispersion of the Verdet coefficients in InSb with different crystal orientations and carrier concentrations are presented in Figure \ref{fig:VerdetResults}. The dispersion data show good agreement with the $\lambda^2$ relation from Equation \ref{eq:Verdet}. The linear dependence of $V$ on $N$ is modified by a linear relationship between $m^*$ and $N$, which approximately accounted for the dependence of the effective mass on the carrier concentration of our samples. This variance of the effective mass with carrier concentration prompted studies to optimize the carrier concentration to produce a maximal FOM \cite{Hilico_2011, Boord_1974}. Accounting for the effective mass dependence on $N$ was the only necessary modification, otherwise the data are well-described within the Drude model.

Using the model from Equation \ref{eq:Verdet}, we extracted the effective mass as a function of the measured carrier concentration. The results of this calculation are shown in Figure \ref{fig:EffMass}. The values calculated here are in good agreement with previous measurements in highly-doped InSb \cite{Spitzer_Fan_1957, Kurnick_Powell_1959}. The effective mass increases at higher carrier concentrations, indicating a reduced curvature of the conduction band. The increasing carrier concentration pushes the Fermi level further up into the conduction band, filling the lowest conduction band states and populating flatter regions of the non-parabolic conduction band \cite{Boer_Pohl_2018, Kittel_2005, Kane_1957, Jung_1991}. Optical transitions of the free-carriers are sensitive to this reduced curvature and result in a smaller Verdet coefficient at higher concentrations than would be expected in a pure free-electron gas. Although the effective mass is smallest at lower carrier concentrations, the magnitude of the Verdet coefficient is still dominated by the linear scaling of $N$ and increases as $N$ increases. However, the Verdet coefficient increases more slowly as the carrier effective mass increases. Since the Verdet coefficient does not depend on the mean free time $\tau$ and is insensitive to carrier scattering mechanisms, unlike the absorption coefficient, the Faraday rotation is well-described by the Drude free-electron gas model once the effective mass variation due to the non-parabolic band structure is taken into account \cite{Smith_1959}.

\subsection{Absorption Coefficient \label{sect:Absorb_Results}}
\begin{figure}[hb]
\includegraphics[width=\columnwidth]{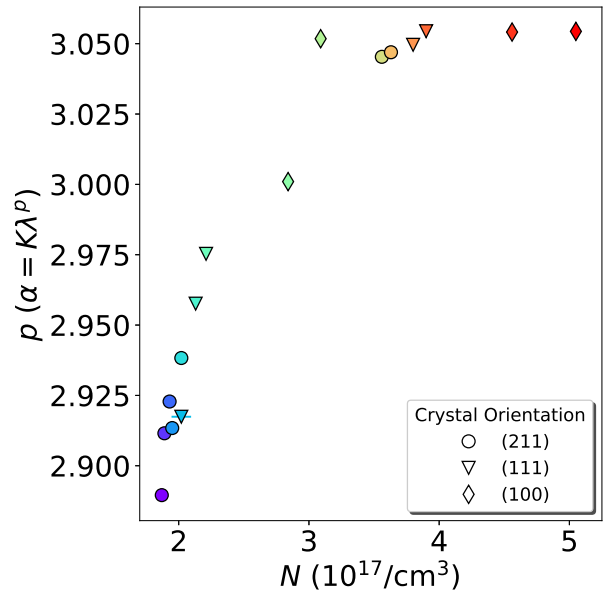}
\caption{The parameter $p$ is plotted for numerical fits of the function $\alpha(\lambda) = K \lambda^p$ to the absorption spectrum of each sample. The marker color scales with the carrier concentration on the horizontal axis. Where they are not visible, error bars have been eclipsed by the point markers. \label{fig:Absorb_Order}}
\end{figure}

\begin{figure*}
\begin{subfigure}{0.49\textwidth}
        \begin{overpic}[unit=1mm,scale=0.49]{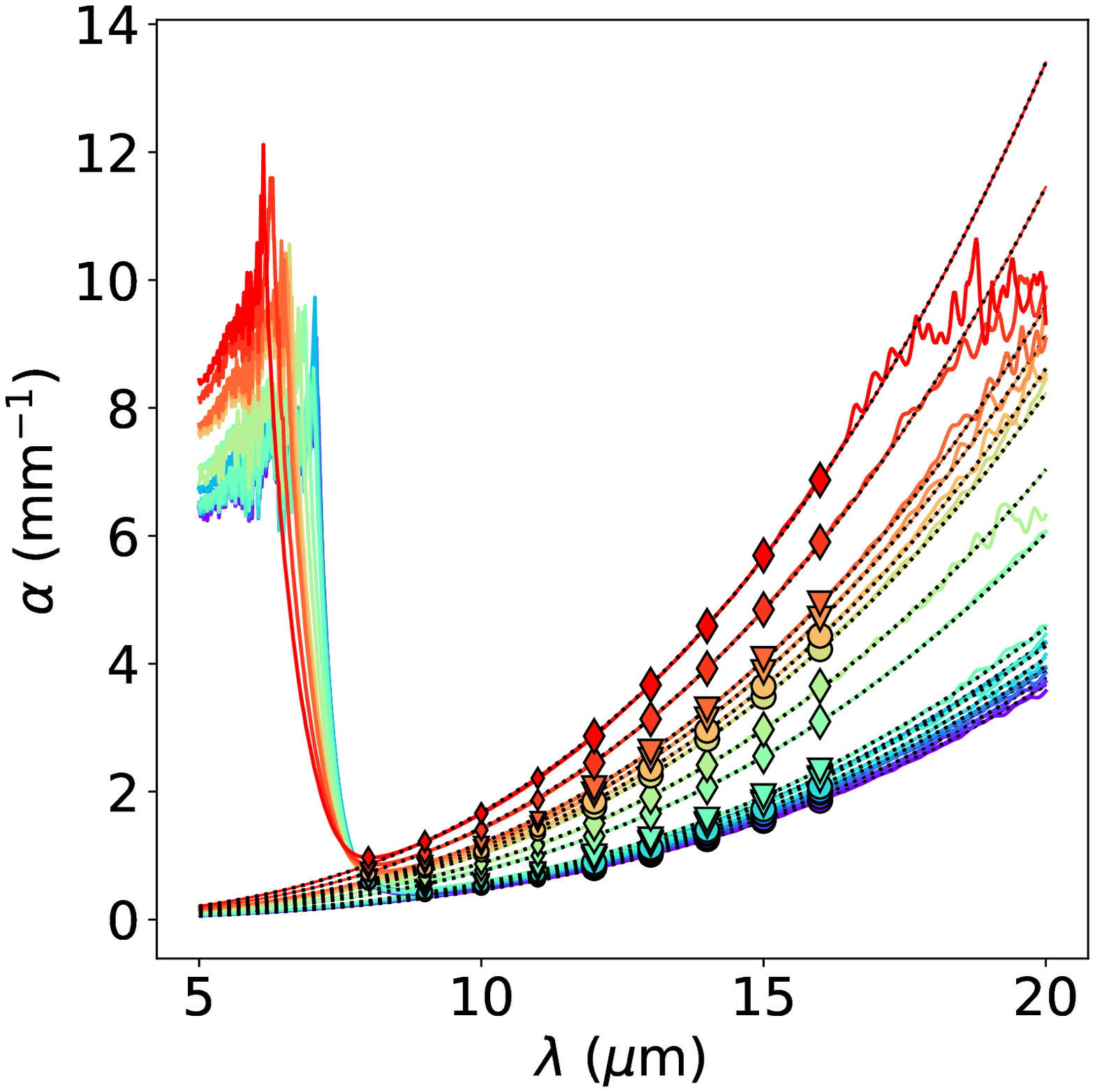}
        \put(84,20){\color{blue}\huge (a)}
        \end{overpic}
        \captionlistentry{}
        \label{fig:Absorb_Lambda}
\end{subfigure}
\hfill
\begin{subfigure}{0.49\textwidth}
        \begin{overpic}[unit=1mm,scale=0.49]{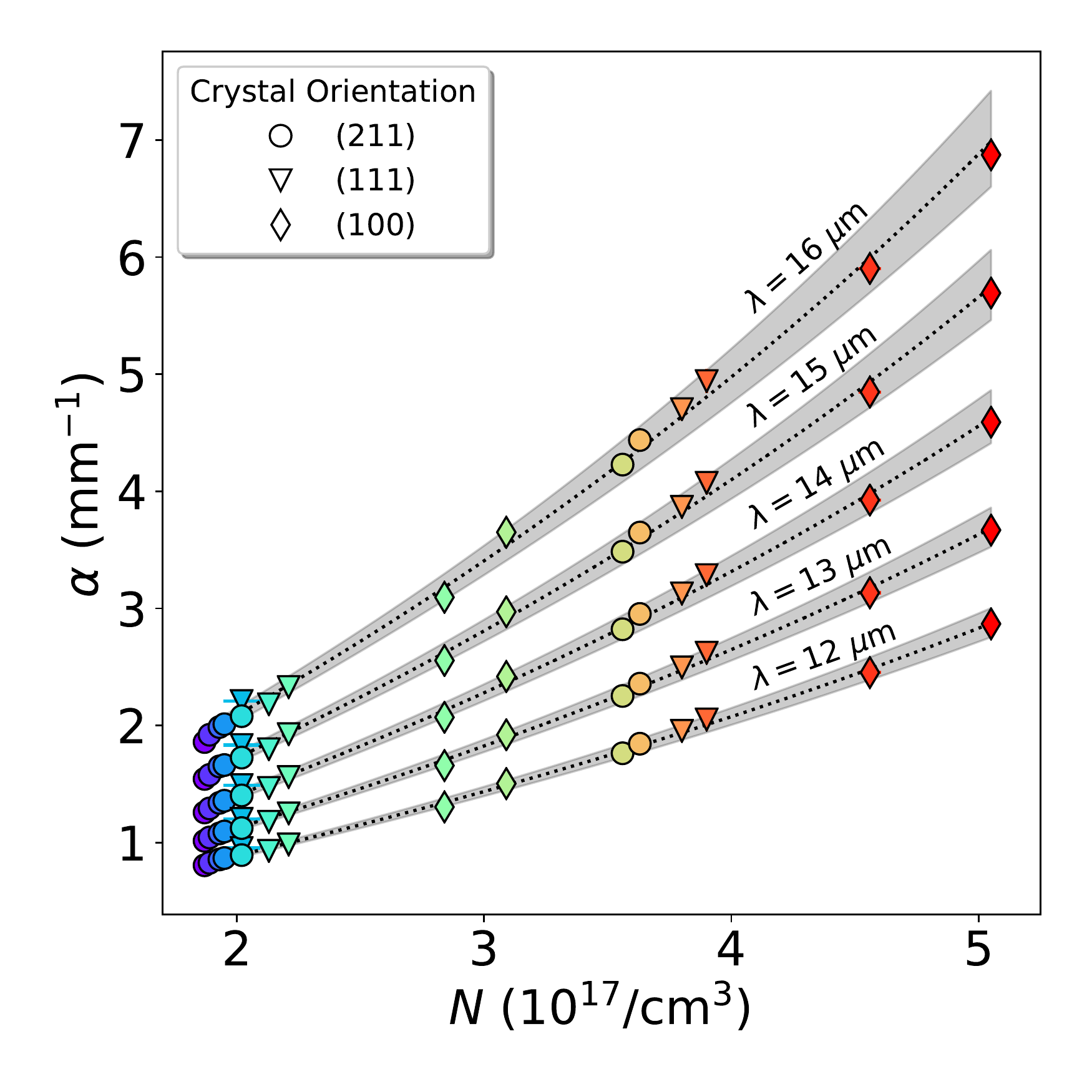}
        \put(84,20){\color{blue}\huge (b)}
        \end{overpic}
        \captionlistentry{}
        \label{fig:Absorb_Doping}
\end{subfigure}
\vspace{-5ex}
\caption{\ref{fig:Absorb_Lambda} Absorption spectra are shown to possess $\lambda^3$ dispersion in the free-carrier regime (actual fitted absorption orders are plotted in Figure \ref{fig:Absorb_Order}). Noise in the spectrum above \SI{6}{\milli\metre^{-1}} at the shortest and longest wavelengths is an artifact of the measurement and was not included in the numerical fit.  \ref{fig:Absorb_Doping} The absorption (at the wavelengths indicated by the markers on the left) is plotted against measured carrier concentration. The numerical fit here is modified by a term in the denominator that implements the linear dependence of effective mass on carrier concentration (results in Figure \ref{fig:EffMass}) in a Drude-like model for absorption. In both figures, the marker color scales with the carrier concentration on the horizontal axis of \ref{fig:Absorb_Doping}. Where the are not visible, error bars have been eclipsed by their point markers. \label{fig:Absorb}}

\vspace*{\floatsep}

\begin{subfigure}{0.49\textwidth}
        \begin{overpic}[unit=1mm,scale=0.49]{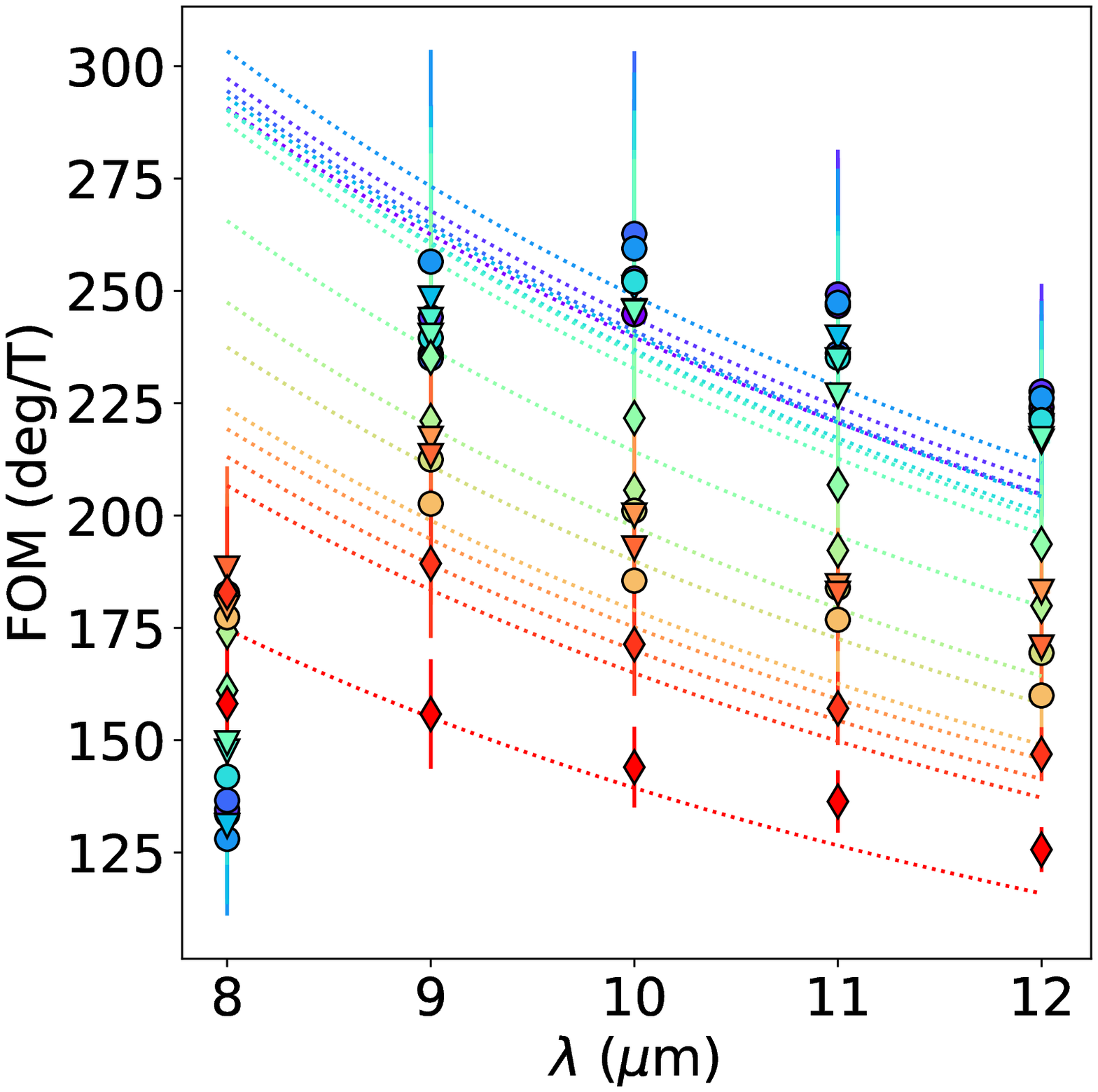}
        \put(85,87){\color{blue}\huge (a)}
        \end{overpic}
        \captionlistentry{}
        \label{fig:FOM_Lambda}
\end{subfigure}
\hfill
\begin{subfigure}{0.49\textwidth}
        \begin{overpic}[unit=1mm,scale=0.49]{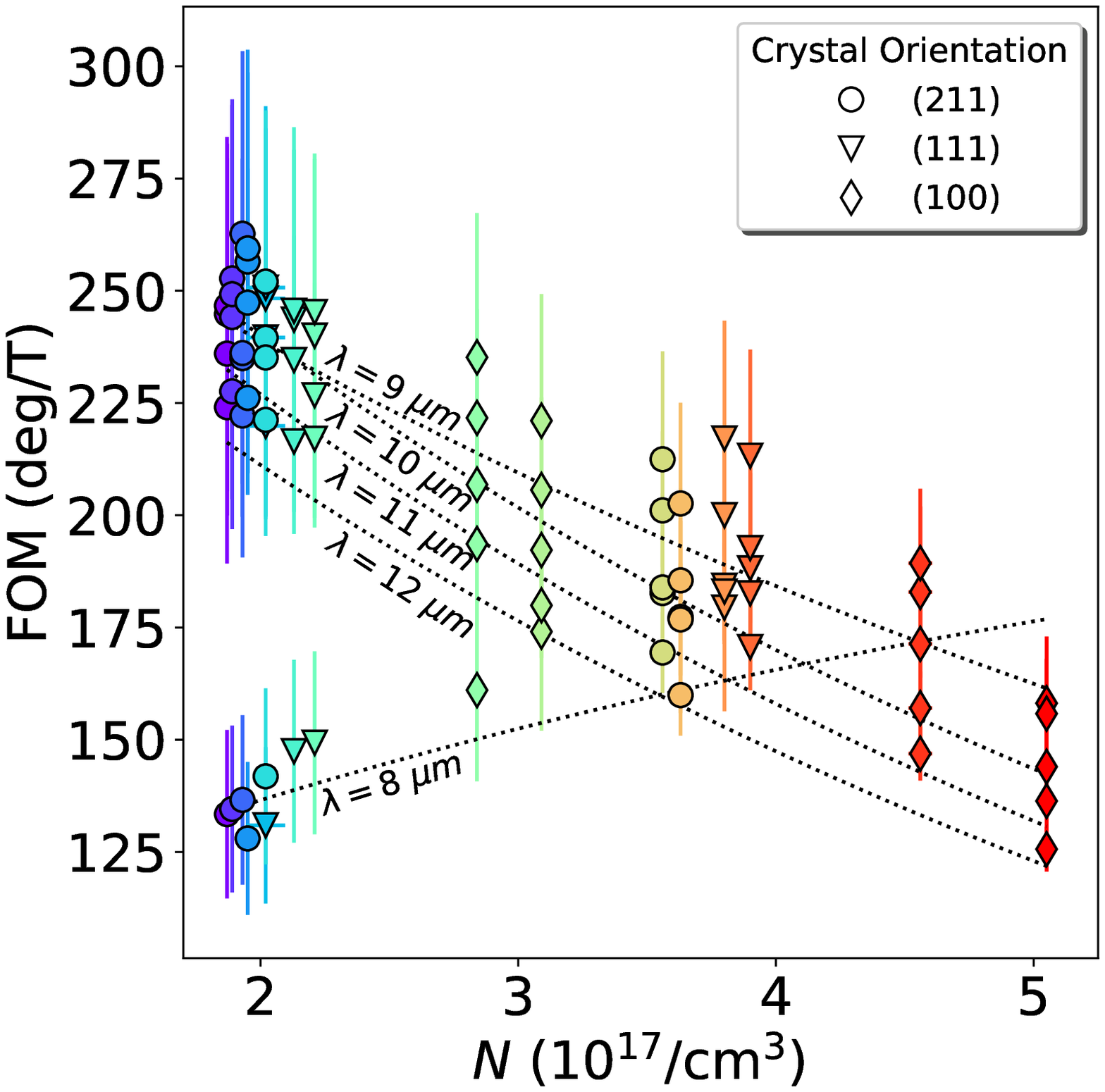}
        \put(54,87){\color{blue}\huge (b)}
        \end{overpic}
        \captionlistentry{}
        \label{fig:FOM_Doping}
\end{subfigure}
\vspace{-5ex}
\caption{The FOM is plotted against wavelength (\ref{fig:FOM_Lambda}) and carrier concentration (\ref{fig:FOM_Doping}). Dotted lines are the FOM as calculated by dividing the absorption and Verdet measurement numerical fits from previous section. It is apparent that the FOM is not isotropic with wavelength and has a $1/\lambda$ dispersion due to the $\lambda^3$ variation of the absorption coefficient. We should therefore expect the FOM to have its maximal value near the optical band edge. Free-carrier absorption models discussed in the text do not account for absorption near the band edge, where the higher absorption coefficient reduced the measured FOM at $\lambda = \SI{8}{\micro\meter}$, especially at lower carrier concentrations where the band gap was narrower due to reduced Moss-Burstein shift. In both figures, marker color scales with the carrier concentration on the horizontal axis of \ref{fig:FOM_Doping}. \label{fig:FOM}}
\end{figure*}

The Drude model predicts the absorption of light in a free-electron gas based on dissipation of energy via scattering mechanisms quantified by a mean free time $\tau$\cite{Jung_1991, Dresselhaus, Boer_Pohl_2018}. Derivations of the absorption coefficient ($\alpha$) based on this model demonstrate a $\lambda^2$ dispersion as shown in Equation \ref{eq:Absorb} \cite{Hilico_2011, Dresselhaus}. Despite the frequent application of this model to InSb, whose abundant free carriers approximate a free-electron gas, it has been shown that this model is not appropriate for the polar semiconductors \cite{Jensen_1971}. A strong interaction between conduction and valence bands in narrow gap materials like InSb forms non-parabolic bands which is at odds with the parabolic energy-momentum relationship of a free-electron \cite{Kane_1957, Basov_1976}. We discussed the effects of the non-parabolic band structure on the effective mass in the previous section. Additionally, the strong spin-orbit splitting in InSb and phonon scattering mechanisms must be included to form an accurate picture of the band structure and energy dissipation. 

Experimental observations have shown, as in our results plotted in Figures \ref{fig:Absorb} and \ref{fig:Absorb_Order}, that the absorption dispersion for InSb actually follows a $\lambda^3$-dependence. This deviation from the Drude $\lambda^2$ law has been attributed primarily to scattering of free-carriers from longitudinal optical phonon modes (polar optical scattering) \cite{Jensen_1971}. Additionally, this suggests the dispersionless FOM in Equation \ref{eq:FOM}, due to matching $\lambda^2$ dispersion of the Verdet and absorption coefficients from the Drude free-electron treatment, requires modification or revision \cite{Hilico_2011, Boord_1974, Smith_1959, Haga_Kimura_1963, Kurnick_Powell_1959}.

Derivations of the free-carrier absorption in InSb based on second-order perturbation theory support the observed $\lambda^3$ dispersion \cite{Haga_Kimura_1963}. The derivation by Jensen provides a general theory of free-carrier absorption in the polar semiconductors, including InSb \cite{Jensen_1973, Jensen_1971, Jensen_1973_summary, Jensen_1979}. Both discuss how the free-carrier absorption in InSb may deviate from $\lambda^2$ and approaches a $\lambda^3$ dispersion due to polar optical or impurity scattering, depending on the specific dopant concentration and crystal quality. If the carrier concentration is roughly equal to the impurity concentration, there exists a crossover where the dominant scattering process transitions from polar optical to ionized impurities around \SI{1e17}{\centi\metre^{-3}} in InSb \cite{Jensen_1973}. Therefore, the smooth transition to sub-$\lambda^3$ absorption below \SI{3e17}{\centi\metre^{-3}} may be a lingering sign of impurity scattering at higher dopant concentrations, as seen in Figure \ref{fig:Absorb_Order} where we have plotted the absorption order $p$ for a presumed absorption law $\alpha = K\lambda^p$ for each carrier concentration. A good test for the presence of impurity scattering would be to measure the absorption at low temperatures where phonons are more likely to be frozen out. Cryogenic cooling may be used in this way to increase the FOM by decreasing the total absorption. %We note, however, that $\lambda^3$ behaviour has been observed in InSb with carrier concentrations as low as \SI{1e16}{\centi\metre^{-3}} where polar scattering dominates over impurity scattering \cite{Jensen_1973}.

\subsection{Figure of Merit}
\begin{figure}[b]
\includegraphics[width=\columnwidth]{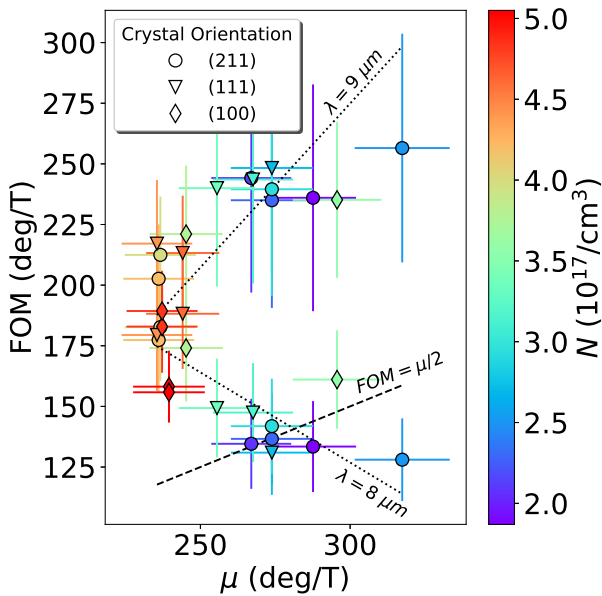}
\caption{The FOM is plotted against mobility for \SI{8}{\micro\metre} and \SI{9}{\micro\metre} wavelengths. A line of best fit is shown for each (short dash) as well as the theoretical line for the FOM from the Drude model (long dash). The \SI{8}{\micro\metre} measurements clearly show the effect of the shifting band edge absorption at lower mobilities (lower carrier concentrations) and have a negative slope. Marker color scales with the carrier concentration as shown in the colorbar at right. Note that $\SI{1}{\centi\metre^2 \volt^{-1} \second^{-1}} = \frac{180}{\pi} \times \SI{e-4}{\tesla^{-1}}$. \label{fig:FOM_Mobility}}
\end{figure}

Our results in Figures \ref{fig:FOM} and \ref{fig:FOM_Mobility} demonstrate that tuning InSb to the optimal FOM for Faraday rotation at a given wavelength, and especially across a range of wavelengths, in the LWIR is not quite as simple as choosing the material with the highest carrier mobility. At the highest mobilities, the FOM of the material was greatly enhanced at longer wavelengths but was even more sharply diminished at shorter wavelengths as a reduced Moss-Burstein shift caused the optical band gap to shrink \cite{Filipchenko_1972, Spitzer_Fan_1957, Kurnick_Powell_1959}. Moreover, the FOM continued to decrease as the wavelength increased away from the optical band edge since the dispersion of the free-carrier absorption coefficient is proportional to $\lambda^3$ while that of the Verdet coefficient is proportional to $\lambda^2$, an effect that was ignored in previous studies which focused on producing isolators at a single wavelength \cite{Hilico_2011, Boord_1974}. As can be seen in Figure \ref{fig:FOM}, the best performance (i.e. maximum polarization rotation with low attenuation) was found at photon energies just below the optical band edge. 

Higher values of the FOM are obtained at the expense of severe attenuation of the shortest wavelengths using carrier concentrations below \SI{2e17}{\centi\metre^{-3}}. Although measurements of the carrier mobility imply that choosing the smallest carrier concentration delivers a large FOM (see Figure \ref{fig:Electric_Mobility}), Figure \ref{fig:FOM} indicates that the ideal carrier concentration must be carefully selected to maximize the FOM across the desired spectral range. For imaging applications across the LWIR spectral range, the data indicate that the ideal InSb should have a carrier concentration greater that \SI{3e17}{\centi\metre^{-3}} where the FOM at \SI{8}{\micro\metre} becomes comparable to the FOM at other wavelengths.

Figure \ref{fig:FOM_Mobility} shows that the Drude FOM (which Equation \ref{eq:FOM} equated to half the mobility as in \cite{Hilico_2011}) is not applicable near the optical band edge energy where we see a negative trend of the FOM with mobility. Even at longer wavelengths, the Drude FOM underestimates the slope of the FOM against mobility. This underestimation is mitigated by the $1/\lambda$ dependence of the true FOM in InSb as the wavelength increases such that a slope of $\frac{1}{2}$ should be expected for one wavelength above the optical band edge and very long wavelengths will have slope $<\frac{1}{2}$. The Drude FOM and measured FOM never match in absolute magnitude in this work; the measured FOM was always nearly double that of the Drude FOM and approximately \SI{70}{\degree \tesla^{-1}} greater than that measured by Hilico et al. in wafers of similar doping near $\lambda = \SI{9}{\micro\metre}$ \cite{Hilico_2011}.

\section{Spectral Filtering via the Faraday Effect \label{sect:Device}}
\begin{figure*}
\begin{subfigure}{0.49\textwidth}
        \begin{overpic}[unit=1mm,scale=0.49]{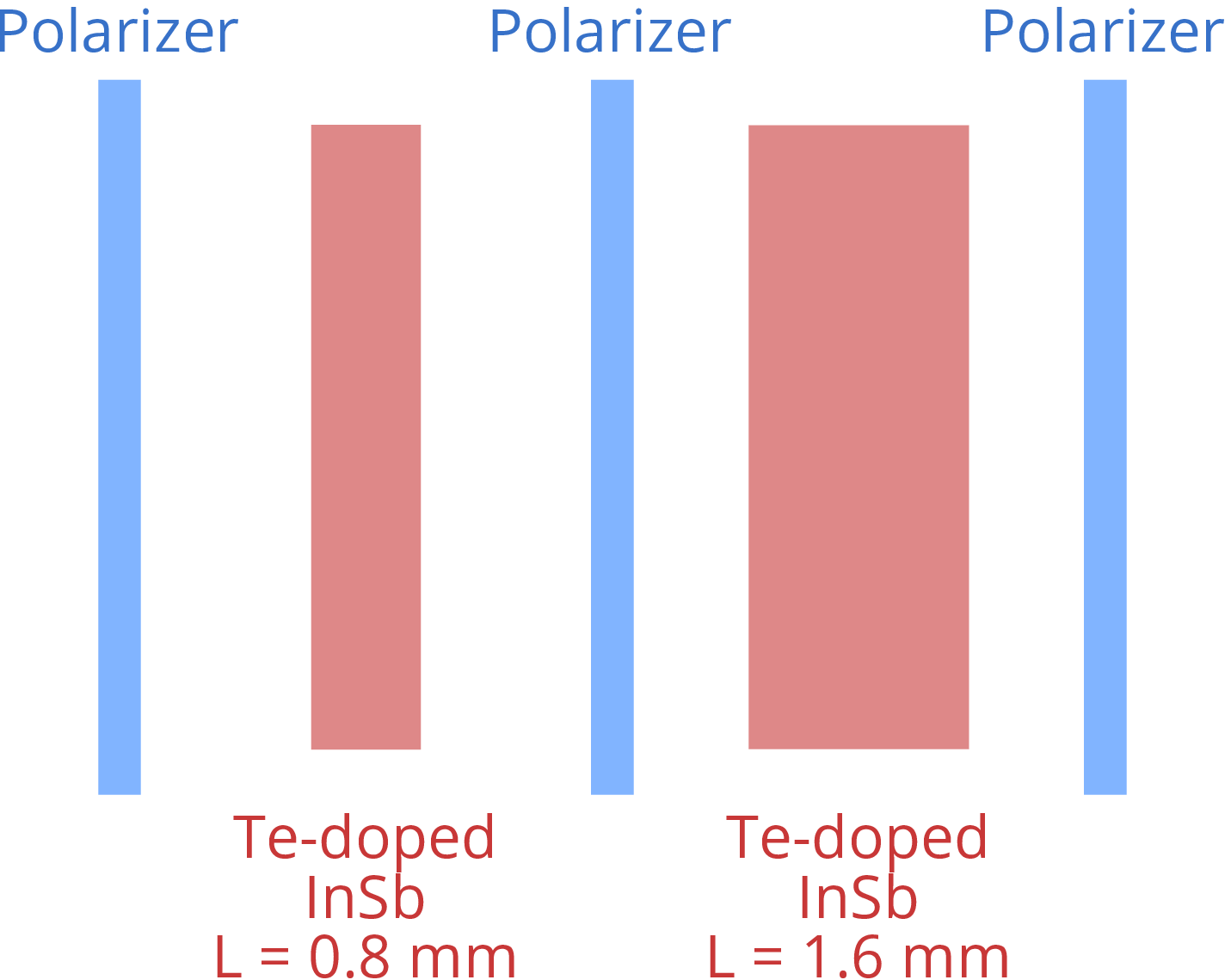}
        \put(0,5){\color{blue}\huge (a)}
        \end{overpic}
        \captionlistentry{}
        \label{fig:Filter_Schematic}
\end{subfigure}
\hfill
\begin{subfigure}{0.49\textwidth}
        \begin{overpic}[unit=1mm,scale=0.49]{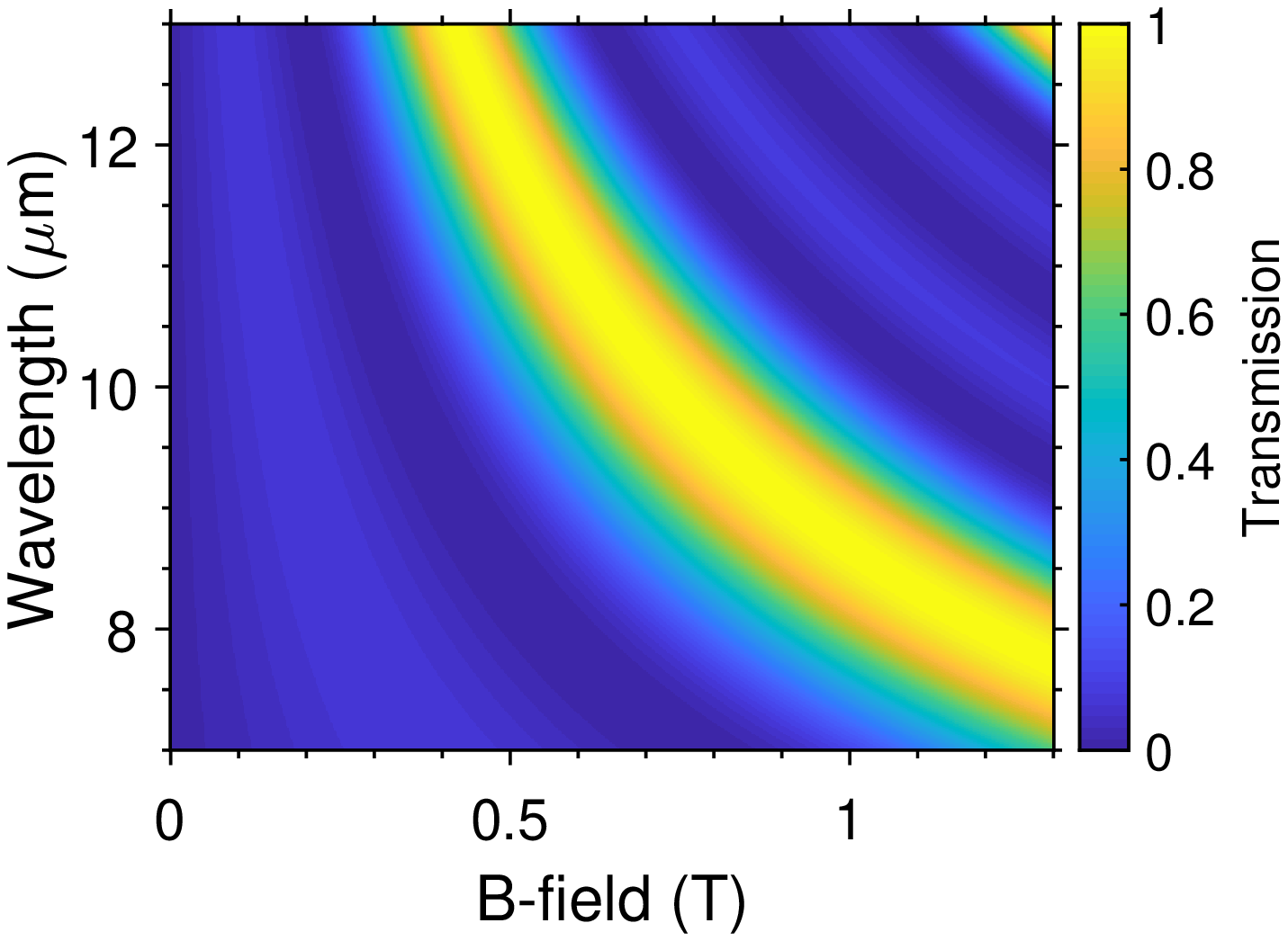}
        \put(15,20){\color{white}\huge (b)}
        \end{overpic}
        \captionlistentry{}
        \label{fig:Filter_Predicted}
\end{subfigure}
\caption{\ref{fig:Filter_Schematic} Schematic of the 2-stage tunable bandpass filter in the demonstration. The Te-doped InSb layers have carrier concentration \SI{2.2e17}{\centi\metre^{-3}} and are anti-reflection coated. The polarizer axes have relative angles of \SI{85.3}{\degree}, \SI{0}{\degree}, and \SI{-170.2}{\degree}. \ref{fig:Filter_Predicted} Theoretical calculation of the filter transmittance normalized to its maximum versus magnetic field strength and wavelength \label{fig:FilterTheory}}
\end{figure*}

Strong dispersion of the Verdet coefficient in highly-doped InSb may be undesirable for constructing a broadband optical isolator, but it can be utilized to assemble a tunable spectral filter \cite{Boord_1974, Hilico_2011}. The basic unit stage of such a  filter has two polarizers positioned at each end of an n-doped InSb. The first polarizer fixes all wavelengths of light from a target to be of a particular polarization; the n-doped InSb rotates the polarization of different wavelengths of light by different amounts; the second polarizer is oriented such that its polarization axis coincides with the rotation of the desired wavelength, thus allowing it to pass to a subsequent stage. The desired wavelength is fully isolated by repeating the basic unit stage and adding additional InSb-polarizer pairs along the optical axis. Figure \ref{fig:Filter_Schematic} shows the schematic of a 2-stage filter; additional stages may be added as needed. Each additional stage will result in further elimination of undesired wavelengths; however, absorption by the InSb and polarizers will eventually limit the number of stages that can be used in practical systems. Spectral filters can be tuned by either varying the magnetic field applied to the doped InSb layers, or by rotating the polarizers while applying a constant or slowly-varying magnetic field to the doped InSb layers.

We experimentally demonstrated a 2-stage tunable bandpass filter as shown in Figure \ref{fig:Filter_Schematic}. Two Te-doped InSb layers with carrier concentration of \SI{2.2e17}{\centi\metre^{-3}} and thicknesses of \SI{0.8}{\milli\metre} and \SI{1.6}{\milli\metre}, respectively, were stacked in between three wire-grid polarizers. The thicker second layer was meant to further narrow down the transmission peak. The three polarizer axes were orientated at relative angles of \SI{85.3}{\degree}, \SI{0}{\degree}, and \SI{-170.2}{\degree}. From the experimentally measured Verdet coefficients of the two InSb layers, the filter transmittance was calculated and is shown in Figure \ref{fig:Filter_Predicted} as a function of magnetic field strength and wavelength. As the field strength is increased from \SI{0.5}{\tesla} to \SI{1.2}{\tesla}, the filter transmission maximum shifts from \SI{12}{\micro\metre} to \SI{8}{\micro\metre}.      

\begin{figure}[b]
\includegraphics[width=\columnwidth]{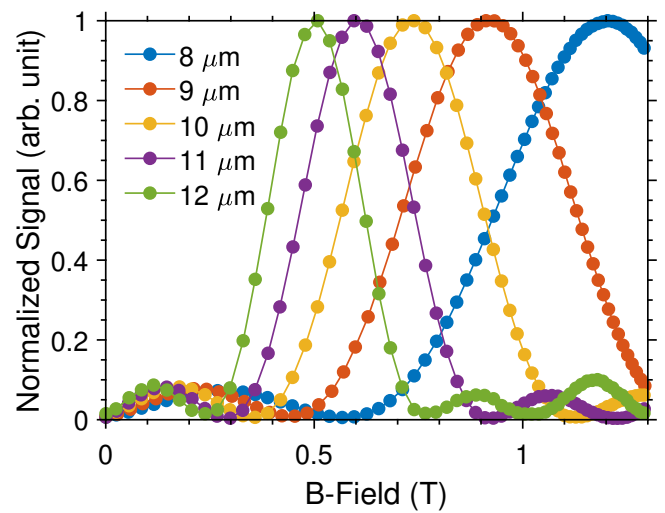}
\caption{Experimental demonstration of the tunable bandpass filter. Transmission at a fixed wavelength between \SIrange{8}{12}{\micro\metre} was measured as a function of the magnetic field strength. As the magnetic field was tuned, the pass band of the 2-stage filter moved between \SIrange{8}{12}{\micro\metre}. \label{fig:FilterExperiment}}
\end{figure}

We mounted the filter on the electromagnet and experimentally demonstrated tunable spectral filtering with the QCL. Figure \ref{fig:FilterExperiment} shows the experimental data of normalized laser transmittance versus magnetic field strength at different wavelengths. The filter tuning behavior agreed well with the theoretical calculation. 

In addition to bandpass filters for hyperspectral imaging applications, it is also possible to design and construct tunable spectral filters using the Faraday effect that block specific wavelengths. Such notch filters may be used to protect detectors from laser jamming and cannot be implemented with tunable filters based on Fabry-Perot etalons \cite{Tuohiniemi_2012, Tuohiniemi_2013}. Additionally, the Faraday effect-based tunable spectral filters provides superior design flexibility and robustness. The lack of precision mechanical parts for tuning optical cavities combined with an ability to create wider pass bands are excellent advantages of our Faraday filter over Fabry-Perot filters.

\section{Conclusion}
We have shown measurements of the absorption and Verdet coefficients in InSb with a range of carrier concentrations and three crystal orientations. These measurements are in agreement with claims in the literature regarding the variation of effective mass with carrier concentration and the nature of the free-carrier absorption in polar semiconductors \cite{Jensen_1973_summary, Jensen_1973, Boord_1974, Smith_1959, Stradling_1970}. We calculated a FOM for the Faraday rotation performance and drew a general conclusion about its dispersion and correspondence to the electrical mobility, namely that a one-to-two relationship cannot generally be drawn between the mobility and the FOM as the Drude model predicts (Equation \ref{eq:FOM}). The dispersion of the FOM, in particular, has gone unnoticed in previous literature. Development of materials for a prototype bandpass filter demonstrated that selection of ideal doping parameters for applications across the LWIR requires additional consideration of competing absorption mechanisms in InSb (optical phonons) and the Moss-Burstein shift of the optical band gap. Our device utilizes the $\lambda^2$ dispersion of the Verdet coefficient to create tunability of the pass band via the magnetic field, laying the foundation for future photonic devices in the LWIR that require wavelength-selective imaging or optical isolation.

\begin{acknowledgments} NP and DC are indebted to Kevin Grossklaus of Tufts University for assistance with Hall effect and ellipsometry measurements as well as insightful discussions. Hall effect measurements and ellipsometry were carried out at the Tufts Epitaxial Core Facility on equipment supported by the United States Office of Naval Research (ONR DURIP N00014-17-1-2591). CAR and JH gratefully acknowledge funding from The Charles Stark Draper Laboratory, Inc.
\end{acknowledgments}

%%%%%%%%%%%%%%%%%%%%%%%%%%%%%%%%%%%%%%%%%%%%%%%%%%%%%%%%%%%%%%%%%%%%%%%%%%%%%
% Place all of the references you used to write this paper in a file
% with the same name as following the \bibliography command
%%%%%%%%%%%%%%%%%%%%%%%%%%%%%%%%%%%%%%%%%%%%%%%%%%%%%%%%%%%%%%%%%%%%%%%%%%%%%

\bibliography{main.bib}
\onecolumngrid

\clearpage

\title{Magneto-Optical Properties of InSb for Infrared Spectral Filtering}
\author{Nolan Peard}
\date{\today}
\affiliation{Department of Applied Physics, Stanford University, Stanford, CA 94205, USA}
\affiliation{The Charles Stark Draper Laboratory, Inc., Cambridge, MA 02139, USA}
\author{Dennis Callahan}
\author{Joy C. Perkinson}
\affiliation{The Charles Stark Draper Laboratory, Inc., Cambridge, MA 02139, USA}
\author{Qingyang Du}
\affiliation{Department of Materials Science and Engineering, Massachusetts Institute of Technology, Cambridge, MA 02139, USA}
\author{Neil S. Patel}
\affiliation{The Charles Stark Draper Laboratory, Inc., Cambridge, MA 02139, USA}
\author{Takian Fakhrul}
\affiliation{Department of Materials Science and Engineering, Massachusetts Institute of Technology, Cambridge, MA 02139, USA}
\author{John LeBlanc}
\affiliation{The Charles Stark Draper Laboratory, Inc., Cambridge, MA 02139, USA}
\author{Caroline A. Ross}
\email{caross@mit.edu}
\author{Juejun Hu}
\email{hujuejun@mit.edu}
\affiliation{Department of Materials Science and Engineering, Massachusetts Institute of Technology, Cambridge, MA 02139, USA}
\author{Christine Y. Wang}
\email{cywang@draper.com}
\affiliation{The Charles Stark Draper Laboratory, Inc., Cambridge, MA 02139, USA}

\onecolumngrid

%Appendix

\section{Supporting Information}

\subsection{Data Analysis}

\begin{figure*}[p]
\begin{subfigure}{0.49\textwidth}
        \begin{overpic}[unit=1mm,scale=0.49]{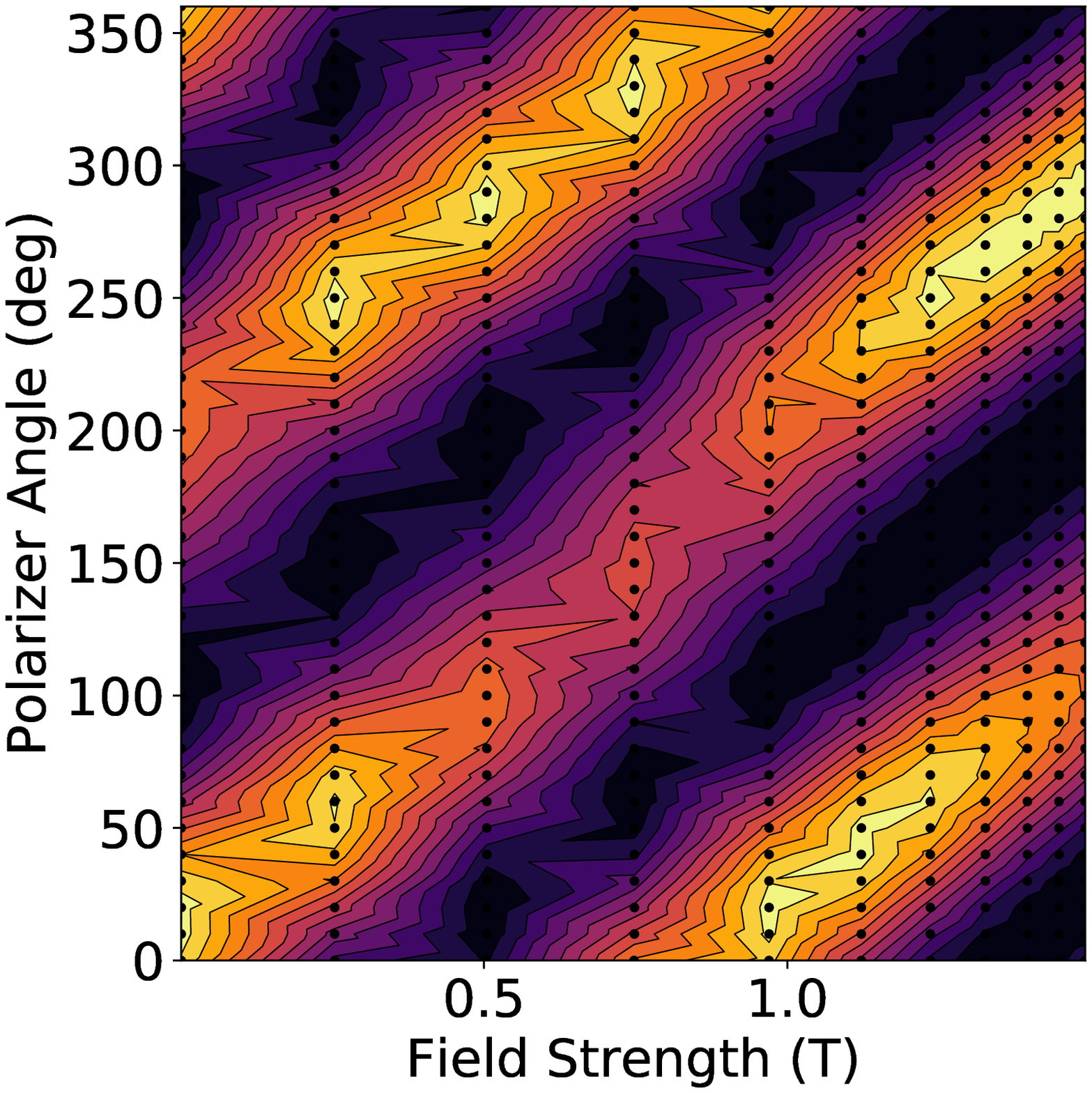}
        \put(80,87){\color{green}\huge (a)}
        \end{overpic}
        \captionlistentry{}
        \label{fig:SI_Fit_Experiment}
\end{subfigure}
\hfill
\begin{subfigure}{0.49\textwidth}
        \begin{overpic}[unit=1mm,scale=0.49]{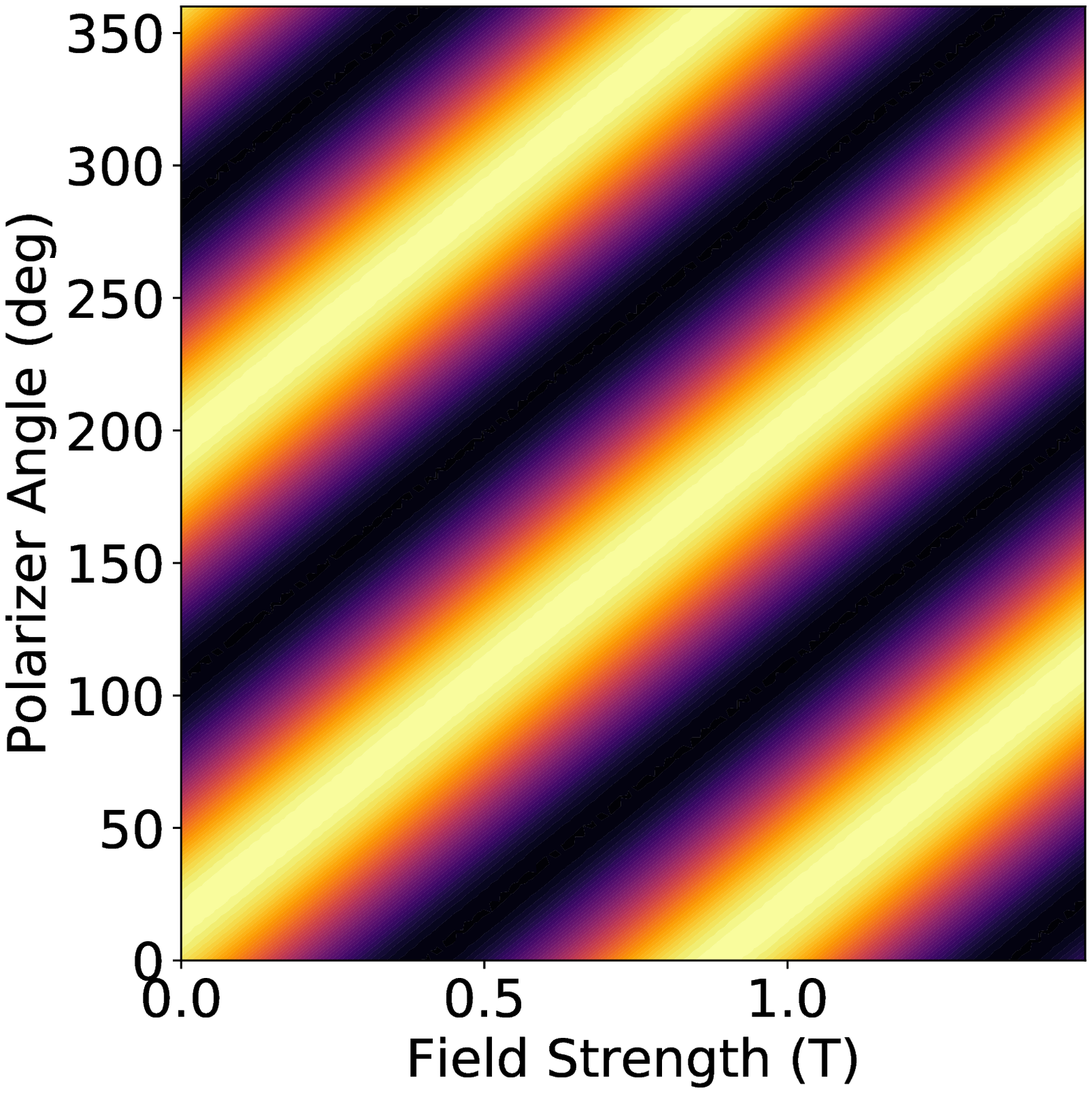}
        \put(80,87){\color{green}\huge (b)}
        \end{overpic}
        \captionlistentry{}
        \label{fig:SI_Fit_Result}
\end{subfigure}
\vspace{-5ex}
\caption{An example of the Verdet measurement and data fitting on an InSb (100) sample with \SI{4.56e17}{\centi\metre^{-3}} carrier concentration at $\lambda = \SI{10}{\micro\metre}$. The sample is \SI{770.8+-6.7}{\micro\meter} thick. \ref{fig:SI_Fit_Experiment} Interpolation of normalized intensity data as a function of magnetic field strength and polarization angle; the solid black dots indicate where the intensity surface was sampled by the experimental setup. The broad attenuation centered around \SI{150}{\degree} is due to polarization bias of the detector. \ref{fig:SI_Fit_Result} Numerical ODR fit to the experimental data used to extract the total polarization rotation (\SI{186.3+-3.14}{\degree \tesla^{-1}}). The measured sample thickness is then used to calculate the Verdet coefficient (\SI{241.6+-2.66}{\degree \tesla^{-1} \milli\metre^{-1}}). \label{fig:SI_Fit}}
\end{figure*}

Figure \ref{fig:SI_Fit} shows an example of fitting detector intensity data via orthogonal distance regression to account for measurement uncertainties in polarizer angle and magnetic field strength.

\subsection{Effective Mass}
\begin{figure*}
\begin{subfigure}{0.49\textwidth}
        \begin{overpic}[unit=1mm,scale=0.49]{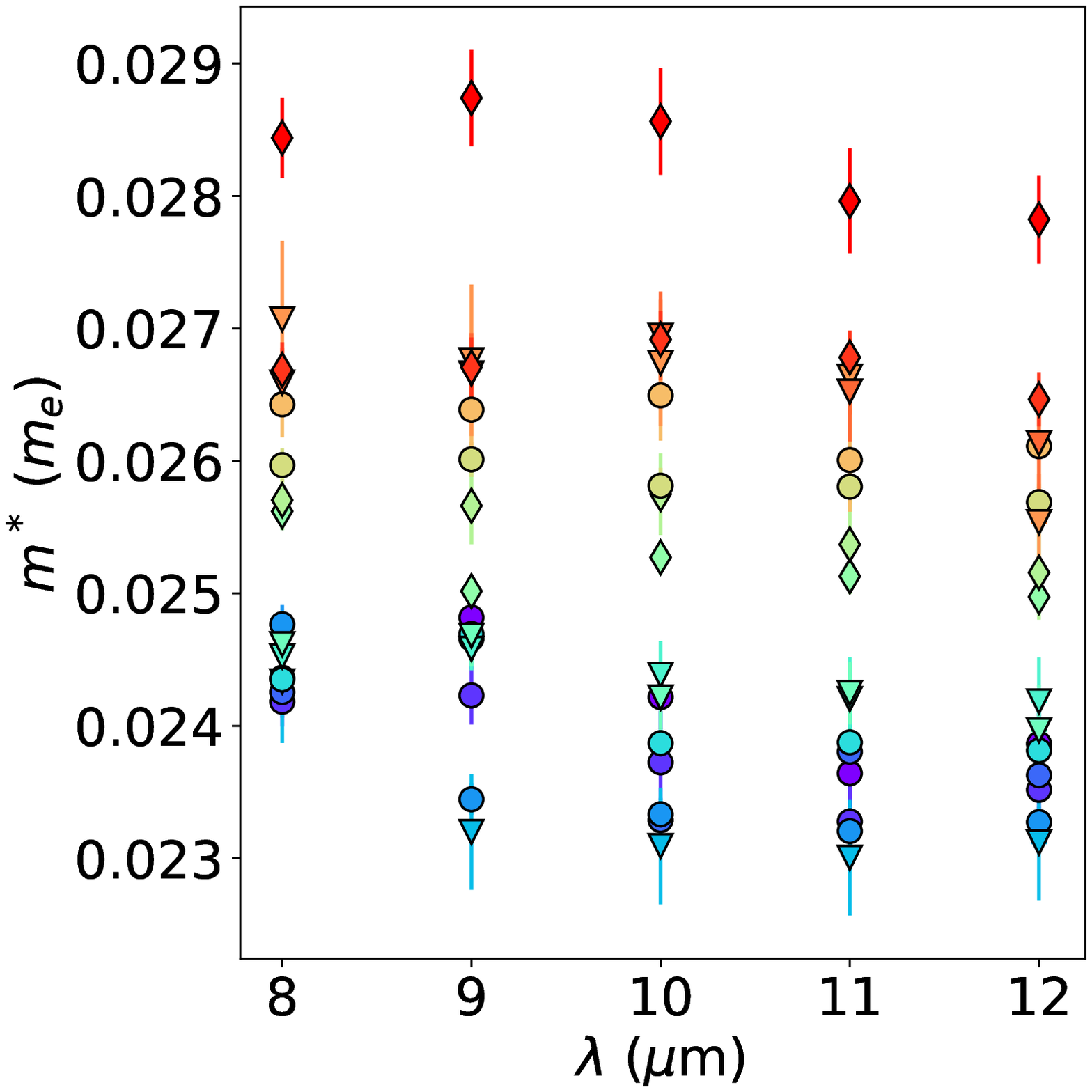}
        \put(85,87){\color{blue}\huge (a)}
        \end{overpic}
        \captionlistentry{}
        \label{fig:SI_EffMass_Lambda}
\end{subfigure}
\hfill
\begin{subfigure}{0.49\textwidth}
        \begin{overpic}[unit=1mm,scale=0.49]{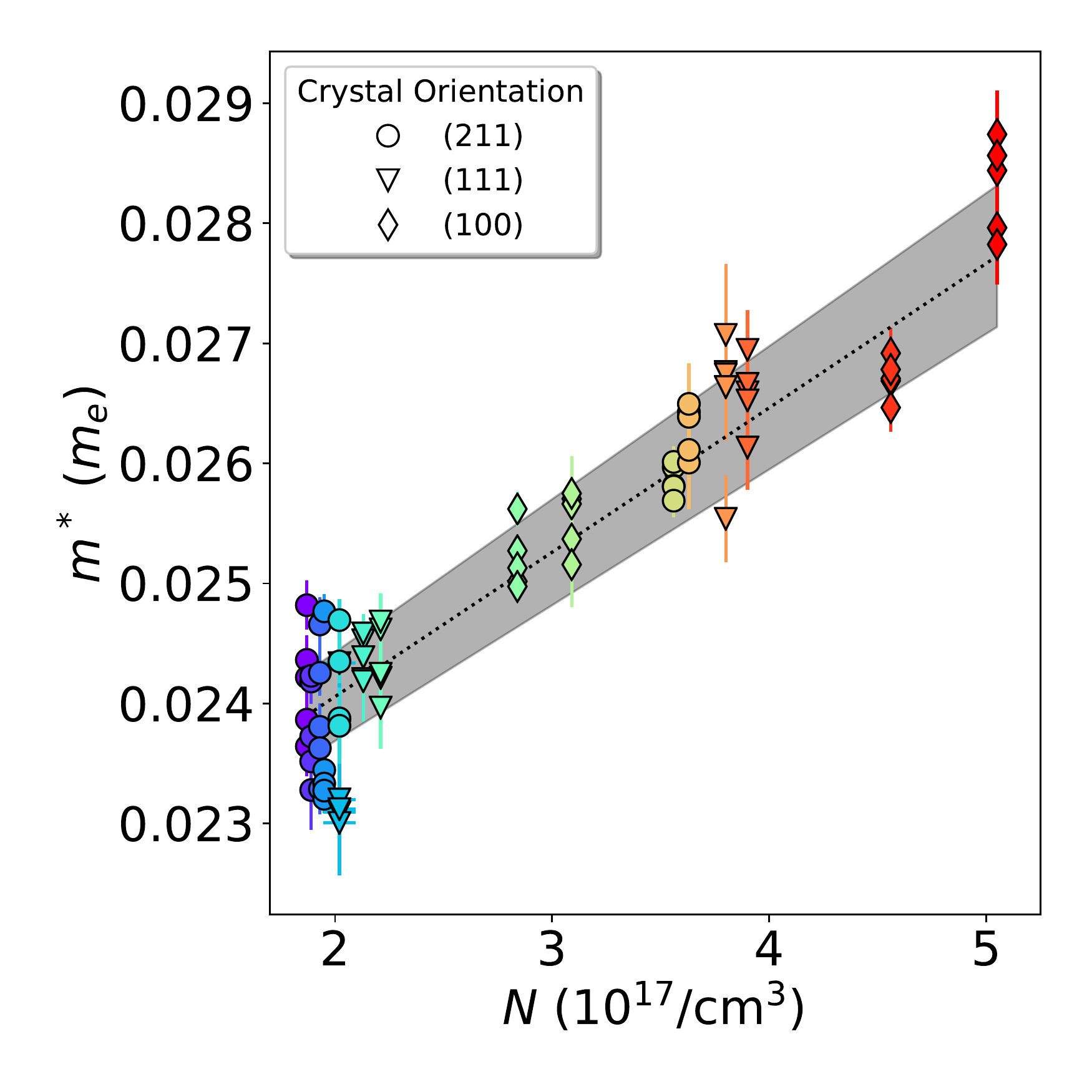}
        \put(84,20){\color{blue}\huge (b)}
        \end{overpic}
        \captionlistentry{}
        \label{fig:SI_EffMass_Doping}
\end{subfigure}
\vspace{-5ex}
\caption{\ref{fig:SI_EffMass_Lambda} shows the variation in the effective mass extracted from the Verdet coefficient measured at different wavelengths. \ref{fig:SI_EffMass_Doping} shows the same information as a function of doping concentration with line of best fit and error region in black. In both figures, the marker color scales with the carrier concentration on the horizontal axis of \ref{fig:SI_EffMass_Doping}.}
\end{figure*}

Figure \ref{fig:SI_EffMass_Lambda} shows the variation in the effective mass with the wavelength of the probe radiation used to measure the Verdet coefficient. Figure \ref{fig:SI_EffMass_Doping} was presented previously in the main text.

% \subsection{Temperature Calibration}
% For measurements of the Verdet coefficient at low temperatures, a Linkam THMS350V temperature controlled stage was used. Care was taken to verify that samples reached their set-point temperature through use of Apiezon N thermal grease and a felt washer which pressed the samples into the thermal grease into close thermal contact with the feedback controlled stage. A DT-670 silicon diode thermocouple from LakeShore Cryotronics was pressed on top of the sample with the felt washer to calibrate the temperature set-point to sample temperature. Multiple samples were measured to develop the calibration curve shown in Figure \ref{fig:Linkam_Temp_Cal}. The thermocouple was subsequently removed for optical measurements in the Verdet measurement setup.  

% \begin{figure}
% \includegraphics[width=\columnwidth]{figures/Linkam_Temp_Calibration_Curve.png}
% \caption{Temperature calibration curves for samples in the Linkam THMS350V stage. The upper plot shows process value temperatures measured on the thermocouple as a function of the stage temperature set point along with a line of best fit. The lower plot shows the difference between the process value and set point temperatures along the same horizontal axis.  \label{fig:Linkam_Temp_Cal}}
% \end{figure}

\subsection{Dispersion of the Refractive Index}

\begin{figure}
\includegraphics[width=\columnwidth]{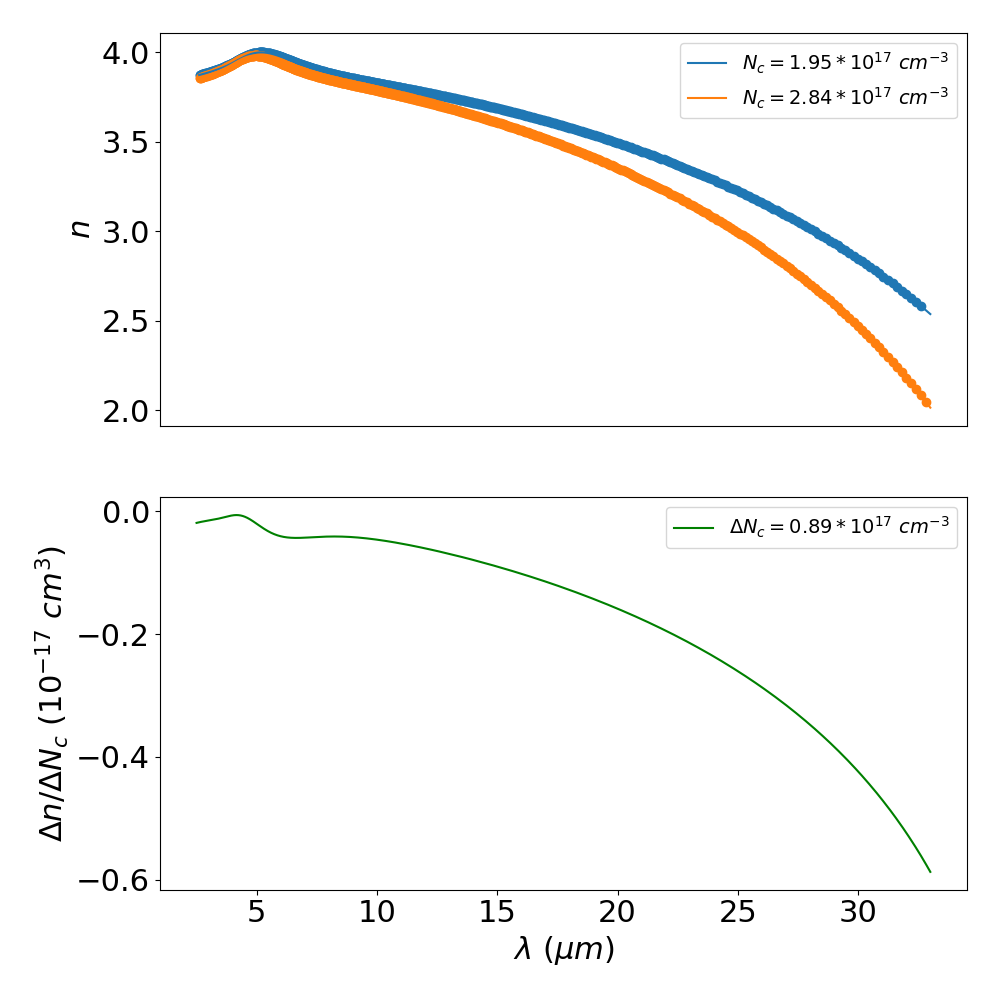}
\caption{Measurements of the refractive index via ellipsometry and the inferred linear variation with carrier concentration of the refractive index at each wavelength. Each sample was measured with identical measurement techniques and the optical constants extracted from identical models, leading to a reliable measurement of $\Delta n(\lambda)$.   \label{fig:SI_n_Disp}}
\end{figure}

\begin{figure*}
\begin{subfigure}{0.49\textwidth}
        \begin{overpic}[unit=1mm,scale=0.49]{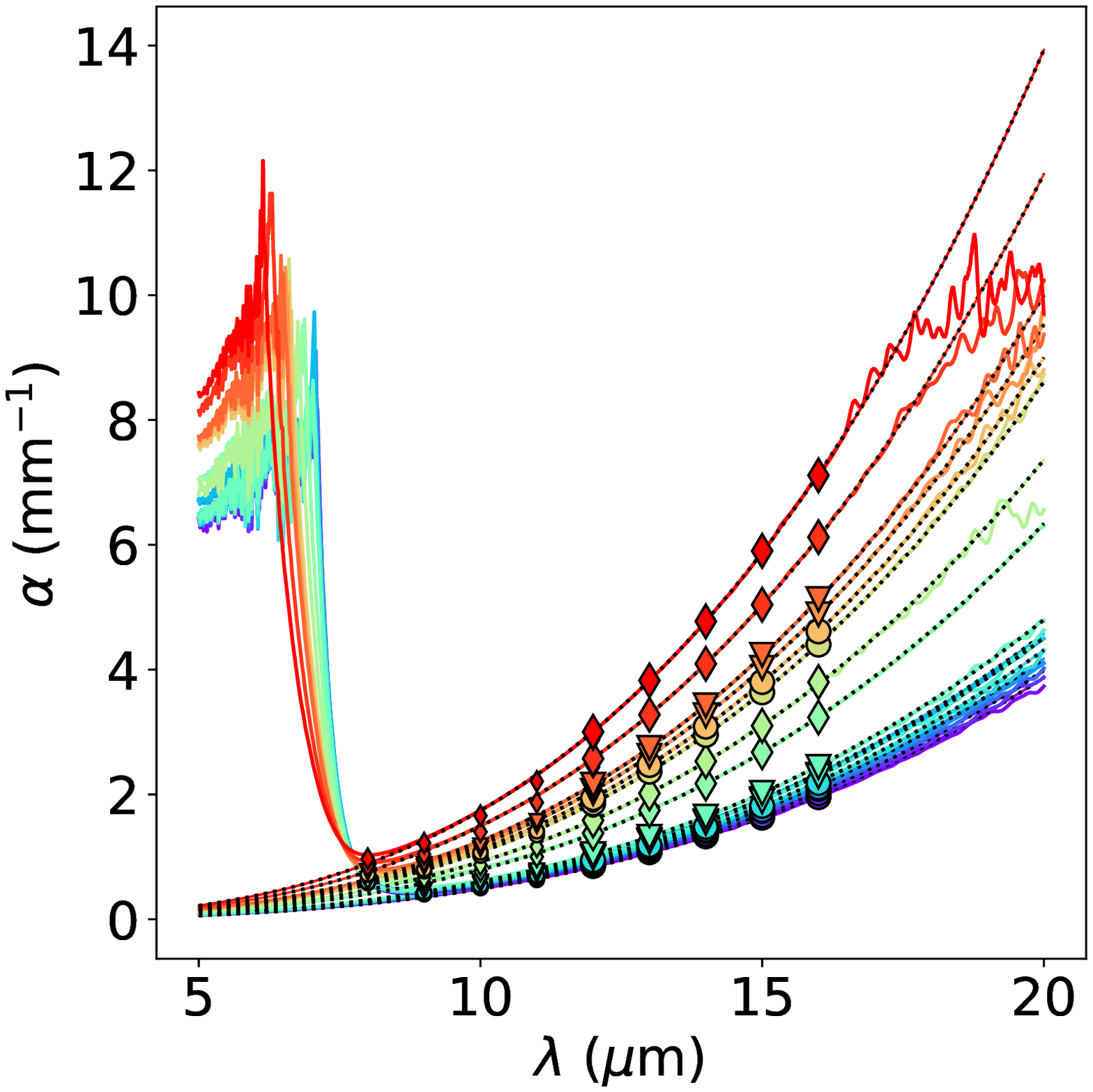}
        \put(84,20){\color{blue}\huge (a)}
        \end{overpic}
        \captionlistentry{}
        \label{fig:SI_Absorb_Lambda}
\end{subfigure}
\hfill
\begin{subfigure}{0.49\textwidth}
        \begin{overpic}[unit=1mm,scale=0.49]{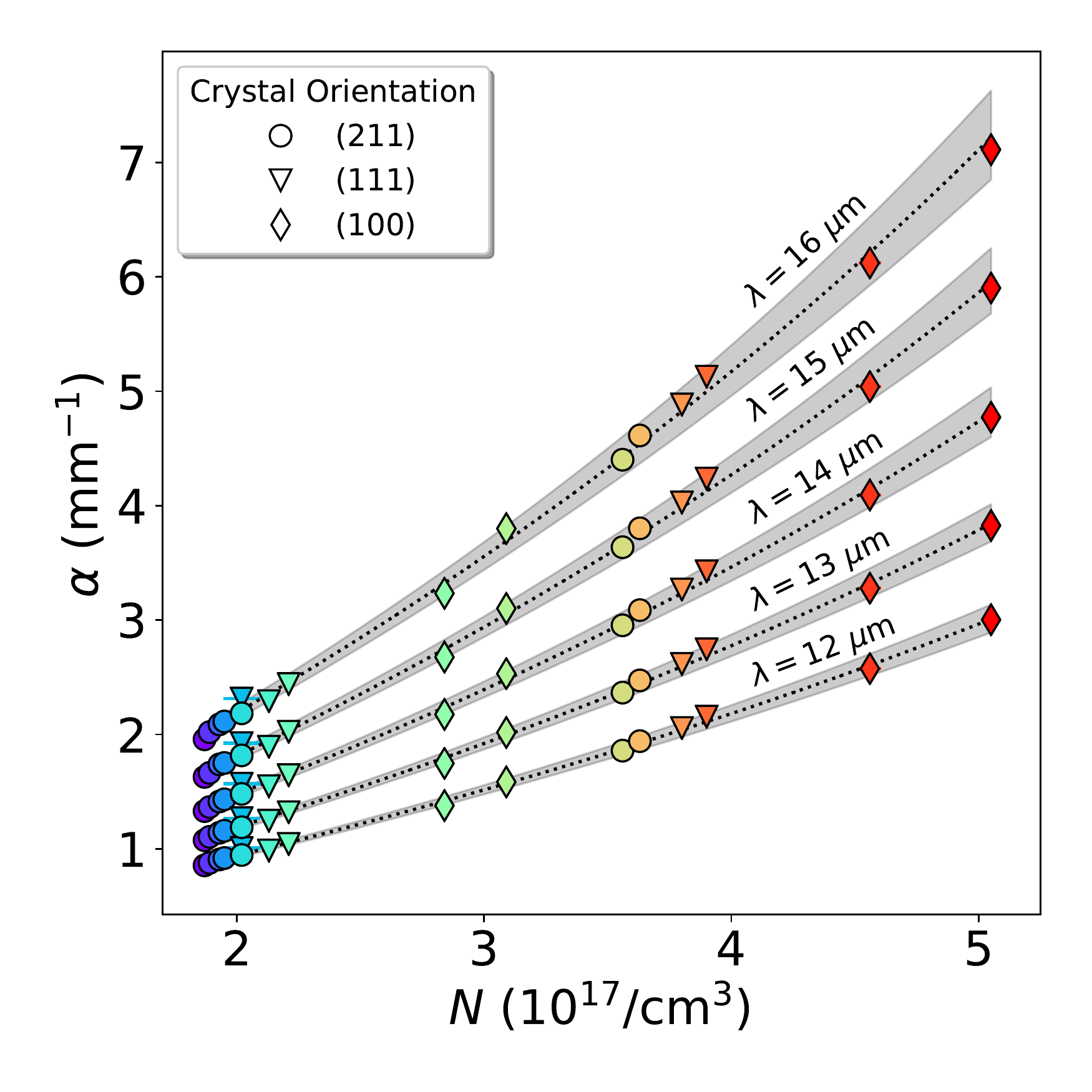}
        \put(84,20){\color{blue}\huge (b)}
        \end{overpic}
        \captionlistentry{}
        \label{fig:SI_Absorb_Doping}
\end{subfigure}
\vspace{-5ex}
\caption{The absorption coefficient $\alpha$ is replotted taking into account the corrections due to the dispersion of the refractive index shown in Figure \ref{fig:SI_n_Disp}. As is evident from the above plots and Figure \ref{fig:SI_Absorb_Order}, the measured absorption coefficient with these corrections shows minimal deviation from the results in the main text which assumed a constant refractive index. In both figures, the marker color scales with the carrier concentration on the horizontal axis of \ref{fig:SI_Absorb_Doping}.} \label{fig:SI_Absorb}
\end{figure*}

We used an IR-VASE ellipsometer to determine whether there is a significant dispersion of the real refractive index, $n$, and to verify that our FTIR absorption measurements are realistic. One side of each sample was roughened to prevent back-reflection and was illuminated with an elliptically polarized light at an incident angle in the vicinity of Brewster angle. The reflected light intensity and phase were subsequently recorded and analyzed to extract the optical constants of the materials. Results are shown in Figures \ref{fig:SI_n_Disp} and \ref{fig:SI_Absorb}.

Although ellipsometry is commonly used to determine the absorption coefficient, we found that the imaginary refractive index of InSb is too small to be measured reliably even with the best ellipsometry methods. The measured signal for absorption falls below the noise floor of our instrument and so measurements of the absolute absorption in the transparent region of InSb between \SIrange{8}{12}{\micro\metre} were not possible with this method. Since it has greater sensitivity, FTIR was used for the absorption measurements in a transmission geometry. However, to calculate the absorption coefficient ($\alpha$), knowledge of the real refractive index is required at all relevant wavelengths. InSb is a high index material ($n \approx 4$) and ellipsometry proved to be a very effective method for measuring $n(\lambda)$. The optical band edge can be determined from the absorption measurements in ellipsometry where the absorption rises above the noise floor of the instrument near the band edge. 

\begin{figure}
\includegraphics[width=\columnwidth]{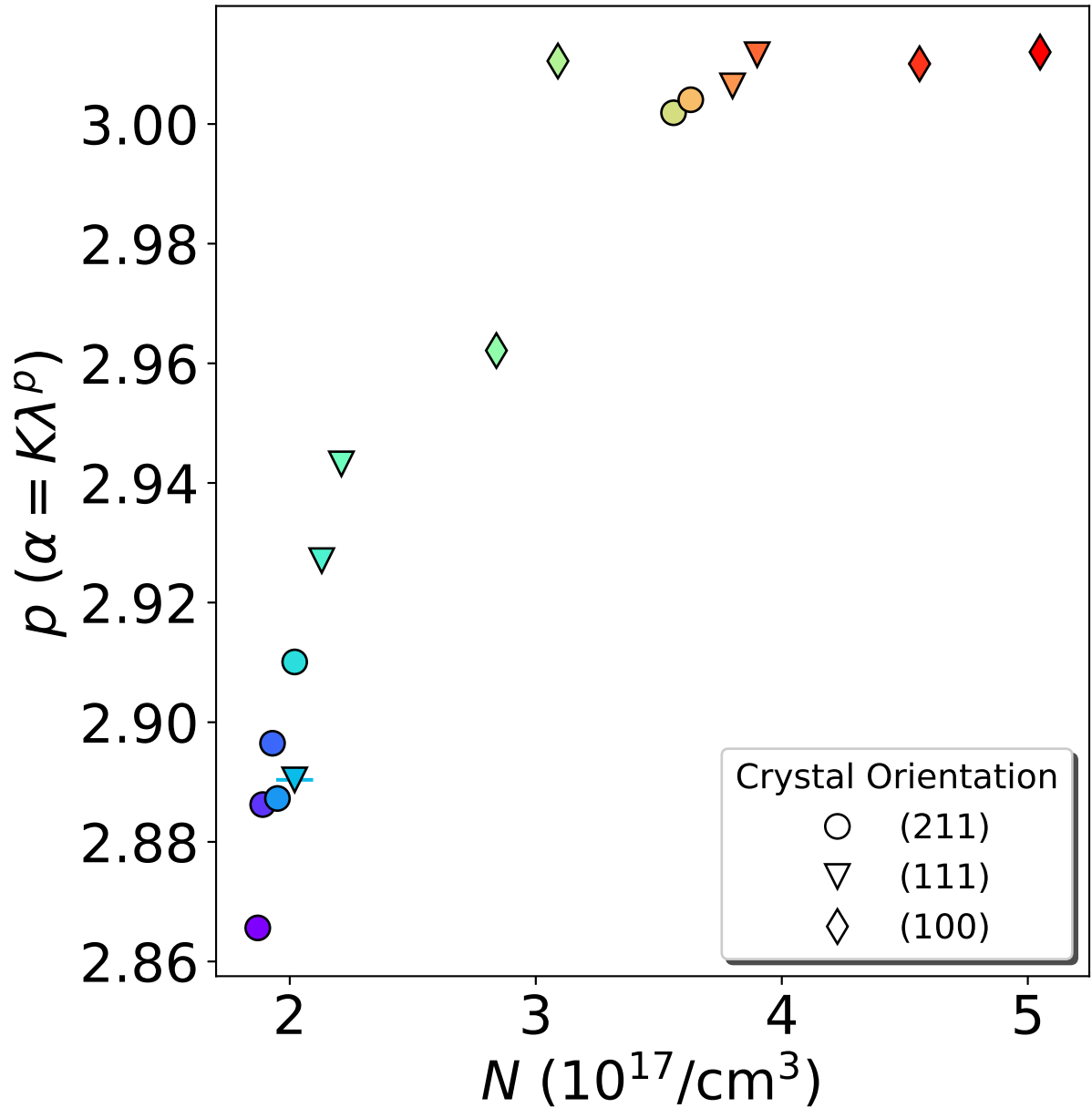}
\caption{The order parameter $p$ is replotted from fits of $\alpha(\lambda) = K \lambda^p$ which include the refractive index dispersion corrections. Again, we see no evidence that the conclusions drawn in the main text are affected by the small corrections for dispersion. The marker color scales with the carrier concentration on the horizontal axis.  \label{fig:SI_Absorb_Order}}
\end{figure}

Using the true $n(\lambda)$ in calculations of the absorption coefficient in Figure \ref{fig:SI_Absorb} resulted in very small changes in the absolute magnitude of $\alpha$ and the order parameter in Figure \ref{fig:SI_Absorb_Order} such that the assumption of a constant $n(\lambda)$ in the main text provided accurate results in the free-carrier regime, despite the use of many samples with different carrier concentrations (and dispersion of $n(\lambda)$). 

\end{document}